\documentclass[a4paper,11pt]{article}
\pdfoutput=1 % if your are submitting a pdflatex (i.e. if you have
\usepackage{jheppub} % for details on the use of the package, please
\usepackage{hyperref}% add hypertext capabilities
\usepackage{booktabs}% more beautiful tables
\usepackage{cleveref}
\usepackage[T1]{fontenc} % if needed
\usepackage{siunitx} % numbers and units
\usepackage{xcolor,graphicx}% Include figure files
\usepackage{dcolumn}% Align table columns on decimal point
\usepackage{bm}% bold math
\usepackage[normalem]{ulem}%Added by Martin, to cross out words with the command \sout{}
\usepackage{braket} % brakets for quantum states 
\usepackage{amsmath}
\usepackage{cancel}%to cross out parts of equations with the command \cancel{}
\usepackage{slashed}
\usepackage{multicol}

\newcommand{\cL}{\mathcal{L}}
\newcommand{\cO}{\mathcal{O}}

\counterwithin*{equation}{section}
\allowdisplaybreaks

\newcommand{\nnl}{\nonumber \\}

\newcommand{\beq}{\begin{equation}}
\newcommand{\eeq}{\end{equation}}
\newcommand{\ba}{\begin{array}}
\newcommand{\ea}{\end{array}}
\newcommand{\bea}{\begin{eqnarray}}
\newcommand{\eea}{\end{eqnarray} }
\newcommand{\be}{\begin{eqnarray}}
\newcommand{\ee}{\end{eqnarray} }
\newcommand{\bal}{\begin{align}}
\newcommand{\eal}{\end{align}}
\newcommand{\bi}{\begin{itemize}}
\newcommand{\ei}{\end{itemize}}
\newcommand{\ben}{\begin{enumerate}}
\newcommand{\een}{\end{enumerate}}
\newcommand{\bc}{\begin{center}}
\newcommand{\ec}{\end{center}}
\newcommand{\bt}{\begin{table}}
\newcommand{\et}{\end{table}}
\newcommand{\btb}{\begin{tabular}}
\newcommand{\etb}{\end{tabular}}

\newcommand{\cM}{{\mathcal M}}

\newcommand{\tev}{\mathrm{TeV}}

\def\hc{{\rm h.c.}}
\usepackage{etoolbox,siunitx}
\robustify\bfseries

\begin{document}

\preprint{\parbox{10cm}{\flushright IFIC/21-16 \\
                                    FTUV/21-0524 \\
                                    CERN-TH-2021-033 \\
                                    MITP-21-013}}

%=======================================================================
\title{EFT at FASER$\nu$}
%=======================================================================

\author[a]{Adam Falkowski,}
\author[b]{Mart\'{i}n Gonz\'{a}lez-Alonso,}
\author[c,d]{Joachim Kopp,}
\author[c,e]{Yotam Soreq,}
\author[f]{Zahra~Tabrizi}

\affiliation[a]{Universit\'{e} Paris-Saclay, CNRS/IN2P3, IJCLab, 91405 Orsay, France}
\affiliation[b]{Departament de F\'isica Te\`orica, IFIC, Universitat de Val\`encia - CSIC, Apt.  Correus 22085, E-46071 Val\`encia, Spain}
\affiliation[c]{Theoretical Physics Department, CERN, Geneva, Switzerland}
\affiliation[d]{PRISMA Cluster of Excellence \& Mainz Institute for Theoretical Physics, Johannes Gutenberg University, Staudingerweg 7, 55099 Mainz, Germany}
\affiliation[e]{Physics Department, Technion—Israel Institute of Technology, Haifa 3200003, Israel}
\affiliation[f]{Center for Neutrino Physics, Department of Physics, Virginia Tech, Blacksburg, VA 24061, USA}

\emailAdd{afalkows017@gmail.com,
          martin.gonzalez@ific.uv.es,
          jkopp@cern.ch,
          soreqy@physics.technion.ac.il,
          ztabrizi@vt.edu}

%=======================================================================
\abstract{
We investigate the sensitivity of the FASER$\nu$ detector to new physics in the form of non-standard neutrino interactions.
FASER$\nu$, which has recently been installed \SI{480}{m} downstream of the ATLAS interaction point, will for the first time study interactions of multi-TeV neutrinos from a controlled source.
Our formalism -- which is applicable to any current and future neutrino experiment -- is based on the Standard Model Effective Theory~(SMEFT) and its counterpart, Weak Effective Field Theory~(WEFT), below the electroweak scale. 
Starting from the WEFT Lagrangian, we compute the coefficients that modify neutrino production in meson decays and detection via deep-inelastic scattering, and we express the new physics effects in terms of modified flavor transition probabilities. 
For some coupling structures, we find that FASER$\nu$ will be able to constrain interactions that are two to three orders of magnitude weaker than Standard Model weak interactions, implying that the experiment will be indirectly probing new physics at the multi-TeV scale. 
In some cases, FASER$\nu$ constraints will become comparable to existing limits -- some of them derived for the first time in this paper -- already with \SI{150}{fb^{-1}} of data.}
%=======================================================================

\maketitle

%=======================================================================
\section{Introduction} 
%=======================================================================

When the CNGS (CERN Neutrinos to Gran Sasso) project ended in 2012, it seemed that, nearly five decades after the first horn-focused neutrino beam had been realized at CERN in 1963~\cite{Dore:2018ldz}, accelerator-based neutrino experiments operating in Europe would become a thing of the past.  
Today, CERN is making a comeback in an unexpected way: with the Forward Search Experiment at the LHC~(FASER)~\cite{Feng:2017uoz,Abreu:2019yak}, 
we are for the first time able to observe neutrinos from a collider experiment~\cite{Abreu:2021hol}.

Of particular interest in this context is FASER$\nu$~\cite{Abreu:2019yak, Abreu:2021hol}, a component of FASER consisting of 1.2~tonnes of tungsten plates, interleaved with thin films of silver bromide emulsion.  
FASER$\nu$ is
installed directly in front of the main detector and, thanks to the emulsion technology, 
offers the superb spatial resolution required especially for reconstructing $\tau$ neutrinos.  
FASER and FASER$\nu$ are located downstream of the ATLAS interaction point at a distance of \SI{480}{m}, and for this reason are
ideal for detecting high-energy neutrinos produced abundantly in the forward direction at the LHC.   
At energies up to several TeV, these neutrinos offer novel opportunities for studying neutrino--nucleon interactions both within the Standard Model~(SM) and beyond.

It is the second possibility -- probing physics beyond the SM -- that we will focus on in this paper. 
Our starting point will be the Weak Effective Theory~(WEFT) Lagrangian, which is the most general effective Lagrangian below the electroweak breaking scale. If non-SM particles are much heavier than the weak scale,  WEFT can be considered a lower-energy descendant of SM Effective Field Theory~(SMEFT)~\cite{Buchmuller:1985jz, Grzadkowski:2010es}, which is a $SU(3)\times SU(2) \times U(1)$ gauge invariant effective theory above the electroweak scale. 
Considering effective operators of dimension-6, we will express the effects of heavy ($\gtrsim \SI{100}{GeV}$) new physics on lower energy neutrino interactions in terms of modified flavor transition probabilities. 
This approach, which was introduced in refs.~\cite{Falkowski:2019xoe,Falkowski:2019kfn}, has the advantage that it is applicable also 
to long-baseline neutrino experiments, where also neutrino oscillations play a role.  
We will then comprehensively study the sensitivity of FASER$\nu$ to a large set of WEFT operators, highlighting the experiment's excellent sensitivity especially to pseudoscalar couplings, to couplings that lead to an anomalous flux of $\tau$ neutrinos, and to anomalous production of charged $\tau$ leptons in $\nu_e$ or $\nu_\mu$ interactions in the detector.

While our results will be derived for the specific case of FASER$\nu$, they will, with minimal modification and rescaling, apply also to other neutrino detectors at the LHC, in particular to the recently approved SND@LHC experiment~\cite{Ahdida:2020evc}.

Searches for non-standard neutrino interactions using LHC neutrinos will add a new chapter to the long-standing search program for such ``NSI'', which has previously focused mostly on neutrino oscillation experiments and other low-energy searches, see for instance refs.~\cite{Bergmann:1999rz, Antusch:2006vwa, Kopp:2007ne, Bolanos:2008km, Ohlsson:2008gx, Delepine:2009am, Biggio:2009nt, Leitner:2011aa, Ohlsson:2012kf, Esmaili:2013fva, Li:2014mlo, Agarwalla:2014bsa, Blennow:2015nxa, Coloma:2017ncl, Farzan:2017xzy, Choudhury:2018xsm, Heeck:2018nzc, Altmannshofer:2018xyo, AristizabalSierra:2018eqm, Esteban:2018ppq,Falkowski:2019xoe}. 
Searches for other types of physics beyond the Standard Model in FASER have previously been discussed in refs.~\cite{Alimena:2019zri, Feng:2017vli, Okada:2019opp, Boiarska:2019vid, Kling:2018wct, Helo:2018qej, Deppisch:2019kvs, Feng:2018pew, Berlin:2018jbm, Dercks:2018eua, Ariga:2018uku, Jodlowski:2019ycu}, while refs.~\cite{Bahraminasr:2020ssz, Kling:2020iar, Bakhti:2020vfq, Bakhti:2020szu} have focused specifically on FASER$\nu$. 
Besides FASER$\nu$ and SND@LHC, additional proposals for detecting LHC neutrinos have been put forwards in refs.~\cite{Buontempo:2018gta, Beni:2019gxv}, highlighting even more the significant community interest in this emerging subfield of neutrino physics.

The paper is organized as follows: 
in \cref{sec:Formalism} we introduce the WEFT formalism and we compute analytically the modified neutrino event
rates in the presence of new dimension-6 interactions in the framework of WEFT. 
In \cref{sec:FASER} we describe our numerical evaluation of these rates as well as the statistical procedure we follow to determine the sensitivity of FASER$\nu$.  
We present our results in \cref{sec:results}, and we compare them with complementary constraints from other experiments in \cref{sec:comparison}. 
We conclude in \cref{sec:conclusion}.

%=======================================================================
\section{Formalism}
\label{sec:Formalism}
%=======================================================================

Let us present in this section the Effective Field Theory~(EFT) formalism that we will use and that was introduced in refs.~\cite{Falkowski:2019xoe,Falkowski:2019kfn}. From an EFT point of view, the most general Lagrangian describing new physics below the electroweak scale is the WEFT Lagrangian, in which the electroweak gauge bosons, the Higgs boson, and the top quark are integrated out while the electroweak symmetry is explicitly broken. 

The part of the WEFT Lagrangian that we will focus on here is the one that FASER$\nu$ will be able to probe.
It is the part that modifies charged-current neutrino interactions with quarks:\footnote{We assume here that total lepton number is conserved.}
\begin{align} 
    \cL_{\rm WEFT} 
    &\supset
    - \,\frac{2 V_{jk}}{v^2} \Big\{
      [ {\bf 1} + \epsilon_L^{jk}]_{\alpha\beta}
             (\bar{u}^j \gamma^\mu P_L d^k) (\bar\ell_\alpha \gamma_\mu P_L \nu_\beta)
    \,+\, [\epsilon_R^{jk}]_{\alpha\beta} (\bar{u}^j \gamma^\mu P_R d^k)
                                          (\bar\ell_\alpha \gamma_\mu P_L \nu_\beta) \nnl
    &\quad +\,
      \frac{1}{2} [\epsilon_S^{jk}]_{\alpha\beta}
             (\bar{u}^j d^k) (\bar\ell_\alpha P_L \nu_\beta)
    - \frac{1}{2} [\epsilon_P^{jk}]_{\alpha\beta}
             (\bar{u}^j \gamma_5 d^k) (\bar\ell_\alpha P_L \nu_\beta) \nnl 
    &\quad +\,
      \frac{1}{4} [\epsilon_T^{jk}]_{\alpha\beta} (\bar{u}^j \sigma^{\mu\nu} P_L d^k)
                                      (\bar\ell_\alpha \sigma_{\mu\nu} P_L \nu_\beta)
    + \hc \Big \} \,.
  \label{eq:EFT_lweft}
\end{align}
Here, $v \equiv (\sqrt{2} G_F)^{-1/2} \approx \SI{246}{GeV}$ is the vacuum expectation value of the SM Higgs field, $V_{jk}$ are the elements of the Cabibbo--Kobayashi--Maskawa~(CKM) matrix, $P_{L,R} = \frac{1}{2} (1 \mp \gamma^5)$ are the chirality projection operators, and $\sigma^{\mu\nu} = \tfrac{i}{2} [\gamma^\mu,\gamma^\nu]$. 
Mass eigenstates of the down-quark, up-quark and charged lepton fields are denoted by $d^k$, $u^j$ and $\ell_\alpha$, respectively.
 The neutrino fields $\nu_\alpha$ are taken in the flavor (weak-interaction) basis here, 
and they are connected to mass eigenstates through the leptonic mixing matrix: $\nu_{\alpha} = \sum_{n=1}^3 U_{\alpha n} \nu_n$. 
We follow the usual convention of denoting flavor indices with Greek letters, and mass eigenstate indices with Roman letters.  
The interaction strengths of the new operators in \cref{eq:EFT_lweft} are parameterized in terms of the dimensionless Wilson coefficients $[\epsilon_X^{jk}]_{\alpha\beta}$, where $j$, $k$ specify the quark generations and $\alpha$, $\beta$ the lepton generations involved in the corresponding dimension-6 operator. 
The index $X$ denotes the Lorentz structure of the operator and can be $X=L,R,S,P,T$ for left-handed, right-handed, scalar, pseudo-scalar, and tensor interactions, respectively. 

In principle, the $[\epsilon_X^{jk}]_{\alpha\beta}$ can be complex; we find, however, that the FASER$\nu$ experiment with which we are mainly concerned in this work, has hardly any sensitivity to their complex phases. 
Therefore, we will treat them as real throughout this paper, with the understanding that the sensitivity estimates we are going to present in \cref{sec:results} apply to the modulus of the $[\epsilon_X^{jk}]_{\alpha\beta}$ in the case of complex coefficients.

Note that we do not consider operators of dimension higher than 6. 
At first sight, this might seem problematic, given that FASER$\nu$ will only be sensitive at quadratic order to some of the dimension-6 operators. 
However, it is not at odds with WEFT power counting.  
Indeed, the observable of interest (the neutrino event rate) will scale as $R/R_\text{SM} = 1 + c \, \epsilon_X^2$, where $R_\text{SM}$ is the count rate without new physics, and $c$ is a numerical factor. 
A dimension-8 WEFT operator with Wilson coefficient $\epsilon_8 / v^4$ will lead to corrections of order $\Delta R/R_\text{SM} \sim \sqrt{c} \, \epsilon_8 E^2/v^2$, where $E$ is the characteristic energy scale of the process, $E^2 \sim E_\nu \cdot \si{GeV}$, in FASER$\nu$.
It is therefore safe to neglect operators of dimension higher than 6 as long as $c \, \epsilon_X^2 \gtrsim \sqrt{c} \, \epsilon_8 E^2/v^2$, or $\sqrt{c} \, \epsilon_X \gtrsim E^2/v^2$ assuming the dimension-6 and higher-order Wilson coefficients are of similar order of magnitude.
We will see from our results in \cref{sec:results} that this condition is always satisfied for models saturating the expected FASER$\nu$ limits.\footnote{The dimension-7 WEFT operators~\cite{Liao:2020zyx} cannot interfere with the SM amplitudes, so their contribution is always smaller than that from dimension-6 operators and can therefore be neglected in this discussion.}

In our analysis, we take the Wilson coefficients in \cref{eq:EFT_lweft} to be given at a renormalization scale $\mu = \SI{2}{GeV}$ in the $\overline{\rm MS}$ scheme. 
This is appropriate for neutrino production in meson decay. 
In principle, neutrino--nucleus interactions in the detector occur at energies about an order of magnitude larger than this, so in principle, one needs to consider renormalization group running in the computation of the neutrino cross-sections. 
Given that RGE effects are expected to be smaller than the systematic uncertainties in our flux predictions, we will neglect them here. 
We will, however, include RG effects when comparing the sensitivity of FASER$\nu$ to the sensitivity of collider searches in \cref{sec:comparison}.

The WEFT Wilson coefficients can be matched onto the parameters of the SMEFT at a scale $\mu\sim m_W$,  see e.g.~refs.~\cite{Cirigliano:2012ab,Jenkins:2017jig,Falkowski:2019xoe}. As discussed in ref.~\cite{Falkowski:2019xoe}, this matching exercise shows that all the $\epsilon_X$ Wilson coefficients receive contributions at $\cO(\Lambda^{-2})$, where $\Lambda$ is the SMEFT ultraviolet completion scale.  
Moreover, one finds $[\epsilon_R^{jk}]_{\alpha\beta} \propto \delta_{\alpha\beta}$ at $\cO(\Lambda^{-2})$, while the remaining $\epsilon_X$ can be off-diagonal in the lepton flavor indices at this order.
However in this paper we allow for flavor-off-diagonal $[\epsilon_R^{jk}]_{\alpha\beta}$, i.e. $\alpha\ne\beta$, which can be generated by new physics below the electroweak scale.

The new interactions introduced in the Lagrangian of \cref{eq:EFT_lweft} can affect the production and detection of neutrinos, modifying the observed event rate, flavor composition, and oscillation pattern compared to the SM.  
Let us assume each neutrino is produced from a source particle $S$ together with a charged lepton $\ell_\alpha$: $S \to X \ell_\alpha \nu_n$.
When this process is mediated at the parton level by the $u^j \bar d^k \to \ell_\alpha^+ \nu_n$ transition (or another transition related by crossing symmetry), the corresponding production amplitude is denoted by $\cM_{\alpha n}^{S,jk}$.
For instance, neutrino production in charged pion decay ($\pi^+ \to \mu^+ \nu_n$) corresponds to a matrix element of the form $\cM_{\mu n}^{\pi,ud}$.
Similarly, we denote by $\cM_{\beta n}^{D,jk}$ the amplitude for neutrino charged-current scattering on a quark in the target nucleus:  $\nu_n d^k \to u^j \ell_\beta^-$. The amplitude for the scattering on an anti-quark, 
$\nu_n \bar u^j \to \bar d^k \ell_\beta^-$, 
is denoted by $\cM_{\beta n}^{\bar D,jk}$.  
The structure of the EFT Lagrangian in \cref{eq:EFT_lweft} implies that 
these amplitudes take the general form
\begin{align}
  \cM_{\alpha n}^{S,jk} &= U_{\alpha n}^{*} A^{S,jk}_{L,\alpha}
    + \sum_{X=L,R,S,P,T} [\epsilon_X^{jk} U]_{\alpha n}^* A^{S,jk}_{X,\alpha} \,, \nnl 
  \cM_{\beta n}^{D,jk}  &= U_{\beta n} A^{D,jk}_{L,\beta}
    + \sum_{X=L,R,S,P,T} [\epsilon_X^{jk} U]_{\beta n} A^{D,jk}_{X,\beta} \,,
\label{eq:Mdecomposition}
\end{align}
and an analogous relation for $\cM_{\beta n}^{\bar{D},jk}$ and $A^{\bar{D},jk}_{X,\beta}$.
These equations implicitly define the reduced matrix elements $A^{S,jk}_{X,\alpha}$ and $A^{D,jk}_{X,\beta}$ (with all PMNS matrix elements and WEFT coefficients factored out). 
For anti-neutrinos \cref{eq:Mdecomposition} holds with $U \to U^*$, and with $A^{S,jk}_{X,\alpha}$, $A^{D,jk}_{X,\beta}$ replaced by the corresponding anti-neutrino amplitudes. 
The differential event rate for neutrinos of flavor $\beta$ at a neutrino detector at a distance $L$ from the source is~\cite{Falkowski:2019kfn}\footnote{%
Ref.~\cite{Falkowski:2019kfn} considered a single monochromatic source and a single type of target particles. In this paper we generalize their expressions to the case of multiple sources moving with different energies in the lab frame, and to account for DIS scattering on the target nucleus. }
\begin{align}
  \frac{dR_\beta}{dE_\nu}
    &= N_T \, \sigma^{\text{SM}}_\beta(E_\nu) \,
       \sum_{\alpha, S} \Phi^{S,\text{SM}}_\alpha(E_\nu) \, 
       \tilde{P}^S_{\alpha\beta}(E_\nu,L) \,,
  \label{eq:masterRate}
\end{align}
where $N_T$ is the number of target particles, $\Phi^{S,\text{SM}}_\alpha(E_\nu)$ and $\sigma^\text{SM}_\beta(E_\nu)$ are the SM neutrino flux from the source $S$ and the SM detection cross section, respectively, both at neutrino energy $E_\nu$ in the target's rest frame. For the anti-neutrino flux and cross-section we use the notation $\bar{\Phi}^{S,\text{SM}}_\alpha$ and $\bar{\sigma}^\text{SM}_\beta$. 
The sum in \cref{eq:masterRate} runs over the flavor of the charged lepton produced together with the neutrino, and over all types of source particles $S$, in particular $S = \pi^\pm, K^0_S, K^0_L, K^\pm$, etc. 
We stress that both $\Phi^{S,\text{SM}}_\alpha(E_\nu)$ and $\sigma^\text{SM}_\beta(E_\nu)$ are calculated in the absence of any new physics. 
The latter is included in the modified flavor transition probabilities given by
{\small
\begin{align}
  \tilde P^S_{\alpha\beta} (E_\nu,L)
    &=      \sum_{n,m} e^{-i \Delta m_{nm}^2 L / (2 E_\nu)} \nnl
    &\hspace{-1.5cm}
     \times \bigg[ U_{\alpha n}^* U_{\alpha m}
          +\! \sum_{X,j,k} p_{XL,\alpha}^{S,jk} [\epsilon_X^{jk} U]_{\alpha n}^* U_{\alpha m}
          +\!\!  \sum_{X,j,k} p_{XL,\alpha}^{S,jk*} U_{\alpha n}^* [\epsilon_X^{jk} U]_{\alpha m}
          +\!\!\! \sum_{X,Y,j,k} p_{XY,\alpha}^{S,jk} [\epsilon_X^{jk} U]_{\alpha n}^*
                                                [\epsilon_Y^{jk} U]_{\alpha m} \bigg] \nnl
    &\hspace{-1.5cm}
     \times \bigg[ U_{\beta n} U_{\beta m}^*
          +\! \sum_{X,r,s} d_{XL,\beta}^{rs}  [\epsilon_X^{rs} U]_{\beta n} U_{\beta m}^*
          +\! \sum_{X,r,s} d_{XL,\beta}^{rs*} U_{\beta n} [\epsilon_X^{rs} U]^{*}_{\beta m}
          +\! \sum_{X,Y,r,s} d_{XY,\beta}^{rs} [\epsilon_X^{rs} U]_{\beta n}
                                             [\epsilon_Y^{rs} U]^{*}_{\beta m} \bigg],
  \label{eq:tildeP}
\end{align}}
where $\Delta m^2_{nm} \equiv m_{\nu_n}^2-m_{\nu_m}^2$ are the neutrino mass squared differences, $L$ is the baseline (the source--detector distance), and $U$ is again the $3 \times 3$ leptonic mixing matrix (the PMNS matrix). 
It is important to note that the expression in \cref{eq:tildeP} is not an oscillation probability in the usual sense. 
In particular, it can be larger than one. 
We use the tilde in $\tilde P^S_{\alpha\beta}$ to remind ourselves of this fact. 
Nevertheless, $\tilde P^S_{\alpha\beta}$ is a useful quantity  as it 
encapsulates all the new physics effects in neutrino production, detection, and propagation in a single quantity. 
In the SM limit, where the Wilson coefficients $\epsilon_X$ are zero, $\tilde P^S_{\alpha\beta}$ reduces to the standard oscillation probability.

The production and detection coefficients, $p_{XY,\alpha}^{S,jk}$ and $d_{XY,\beta}^{jk}$, quantify how strongly a given WEFT operator affects the neutrino event rate.
Roughly speaking,
the production coefficients parameterize
the ratio of the decay width into neutrinos of energy $E_\nu$ including new physics to the decay width into neutrinos at the same energy in the SM. 
Similarly, the detection coefficients parameterize
the ratio between the interaction cross section for neutrinos of energy $E_\nu$ including new physics and the one without new physics.
They carry indices indicating the Lorentz structure of the interaction ($X$, $Y$), the charged lepton flavor ($\alpha$, $\beta$), and the involved quark flavors ($j$, $k$).

The production coefficients are  defined as
\begin{align}
    p_{XY,\alpha}^{S,jk} &\equiv
      \frac{\int dE_S \frac{\phi_S(E_S)}{E_S}
            \sum_i \beta^S_i(E_S) \int\! d\Pi_{P'_i} A_{X,\alpha}^{S_i,jk} 
                                      A_{Y,\alpha}^{S_i,jk*}}
           {\int dE_S \frac{\phi_S(E_S)}{E_S}
            \sum_{i'j'k'}\beta^S_{i'}(E_S) \int\! d\Pi_{P'_{i'}} |A_{L,\alpha}^{S_i,j'k'}|^2} \,
    \label{eq:pxy}
\end{align}
Here $dE_S \, \phi_S(E_S)$ denotes the number of source particles of type $S$ with lab frame energy in a small interval $dE_S$ around $E_S$.  
The sum over $i$ enumerates all decay modes of $S$ that at the parton level include the transition $u^j \bar d^k \to \ell_\alpha \nu$. 
For instance, if $S = K^+$, we need to sum over the leptonic and semileptonic decay channels.
The differential $d\Pi_{P'_i}$ in \cref{eq:pxy} is the phase space integration measure for the $i$-th production process, but without the integral over neutrino energy: $d\Pi_{P_i} \equiv  d\Pi_{P'_i} \, dE_\nu$.
Implicitly, this differential also implies sums/averages over polarizations and any other unobserved degrees of freedom. The factor $1/E_S$ that appears in both the numerator and in the denominator of \cref{eq:pxy} arises because the lab frame decay rate of $S$ is proportional to $1/E_S$.
Finally, the factor $\beta^S_i(E_S)$ appearing in the numerator and the denominator of the production coefficients describes the experimental acceptance corresponding to a given production channel $i$. In most cases, it is possible to absorb this factor into the definition of $\phi_S(E_S)$. The one exception relevant to us will be neutrino production in kaon decays, $K_L, K_S, K^\pm$, in which case the acceptance for neutrinos from 2-body decays differs from the one for neutrinos from 3-body decays because of the different energy spectra. We determine each of the corresponding $\beta^S_i(E_S)$ factors as the ratio between the relative flux fraction of channel $i$ at FASER$\nu$ and the SM branching ratio.

The detection coefficients for neutrinos are defined as 
\begin{align}
    d_{XY,\beta}^{jk} 
=   \frac{ \sum_{N=p,n} n_N \int \! dx \, dQ^2 \,
    x^{-2}\big [ A^{D,jk*}_{X,\beta} A^{D,jk}_{Y,\beta} f^N_{q_k} (x, Q^2)+
    A^{\bar D,jk*}_{X,\beta} A^{\bar D,jk}_{Y,\beta} f^N_{\bar q_j} (x, Q^2) \big ] }
    { \sum_{N=p,n} \sum_{j'k'} n_N \int \! dx \, dQ^2 \, x^{-2}\big [ 
    A^{D,j'k'*}_{L,\beta} A^{D,j'k'}_{L,\beta}  f^N_{q_{k'}}(x, Q^2)
    + A^{\bar D,j'k'*}_{L,\beta} A^{\bar D,j'k'}_{L,\beta}  f^N_{\bar q_{j'}}(x, Q^2) \big ]  } 
    \, . 
    \label{eq:dxy}
\end{align}
Here, $n_p$ and $n_n$ are the numbers of protons and neutrons in the target nucleus. 
The Bjorken variable $x$ is  the fraction of the nucleon momentum that is carried by the initial state quark and $Q^2 = - (p_{\ell_\beta}-p_\nu)^2$ is the momentum transfer.
The functions $f^N_q(x, Q^2)$ are the parton distribution functions~(PDFs) describing the quark $q$ content of the nucleon $N$, where we set the factorization scale equal to $Q$. 
For anti-neutrinos, the role of quarks and anti-quarks is swapped in \cref{eq:dxy}, and the corresponding detection coefficients will be denoted by $\bar d_{XY,\beta}^{jk}$. 

Our definition of the production coefficients is valid only if the neutrino flux from a given type of source particle $S$ depends linearly on the decay width. 
This is the case for neutrinos from $\pi^\pm$, $K^0_L$, and $K^\pm$ decays because only a small fraction of these mesons with $E_S \gtrsim 100$~GeV can decay before being stopped in the material surrounding the LHC beam pipe and tunnel. 
In other words, the exponential decay law can be linearized for these decays. 
Our production coefficients are also valid for $K^0_S$ and charm meson decay because for these mesons, the branching ratio into leptonic and semileptonic modes is fairly small so that new physics is only going to affect the branching ratios, but not the total decay rate. \Cref{eq:pxy} would, however, not be valid for a hypothetical parent particle whose lab-frame lifetime is comparable to or smaller than its free-flight distance, and whose total decay width is significantly affected by new physics.

Note that the production and detection coefficients satisfy
\begin{align}
    p_{XY,\alpha}^{S,jk} = [p_{YX,\alpha}^{S,jk}]^*
    \qquad\text{and}\qquad
    d_{XY,\beta}^{rs} = [d_{YX,\beta}^{rs}]^* \,.
\end{align}

A comment is in order on the SM neutrino fluxes $\Phi^{S,\text{SM}}_\alpha(E_\nu)$ appearing in \cref{eq:masterRate}. 
Predicting these fluxes requires simulations, but simulations typically depend on many input parameters that are derived from data, for instance measured decay rates and branching ratios. 
Therefore, in the presence of new physics that affects decay rates and branching ratios, the predictions of typical simulations may be contaminated by new physics effects. 
Such simulations are therefore unsuitable for predicting the purely Standard Model fluxes $\Phi^{S,\text{SM}}_\alpha(E_\nu)$. 
Furthermore, the simulations depend on the CKM matrix elements, which have to be determined experimentally and again may be ``contaminated" by new physics. 
In this study we will ignore this problem because, at this point, the difference between simulated fluxes and those obtained from first principles are much smaller than the uncertainties from other sources, e.g. from imperfect modeling of forward QCD scattering or from nuclear structure effects in neutrino-nucleus scattering.
In a fit to actual data, however, care should be taken so that those pieces of the simulation that are prone to new physics contamination -- namely meson decay rates and branching ratios -- are derived from first principles without input from data. 
This will not lead to an increase of systematic uncertainties because already the calculation of the production coefficients according to \cref{eq:pxy} depends on similar first-principles calculations and therefore suffers from similar systematic uncertainties. 
Furthermore, in these future studies one should adopt some consistent CKM input scheme, for example the one proposed in ref.~\cite{Descotes-Genon:2018foz}, to address the problem of CKM contamination.
In a similar way, also the neutrino interaction cross section entering \cref{eq:masterRate} as $\sigma^{\text{SM}}_\beta(E_\nu)$ should be calculated without using any measurements that may be affected by the WEFT operators we are aiming to constrain. 
This is a non-trivial task, given that the cross section depends on PDFs, and the measurements from which PDFs are extracted can be affected by the effective operators considered in this work. 
See e.g.\ ref.~\cite{Carrazza:2019sec} for a possible approach to disentangling new physics from the PDF fits.

Before we proceed to the calculation of the production and detection coefficients $p_{XY,\alpha}^{S,jk}$ and $d_{XY,\beta}^{rs}$, let us remark that, in the case of FASER$\nu$ with its very high neutrino energies and rather short baseline, it is justified to set $L=0$, in which case we can simplify \cref{eq:tildeP}:
{\small
\begin{align}
  \tilde P^S_{\alpha\beta} (E_\nu)_{L=0}
    &=      \sum_{\gamma\delta}
     \bigg[ \delta_{\alpha\gamma} \delta_{\alpha\delta}
          + \sum_{X,j,k} p_{XL,\alpha}^{S,jk} [\epsilon_X^{jk}]_{\alpha\gamma}^* \delta_{\alpha\delta}
          + \sum_{X,j,k} p_{XL,\alpha}^{S,jk*} \delta_{\alpha\gamma} [\epsilon_X^{jk}]_{\alpha\delta}
          + \sum_{X,Y,j,k} p_{XY,\alpha}^{S,jk} [\epsilon_X^{jk}]_{\alpha\gamma}^*
                                                [\epsilon_Y^{jk}]_{\alpha\delta} \bigg] \nnl
    &\hspace{0.6cm}
     \times \bigg[ \delta_{\beta\gamma} \delta_{\beta\delta}
          + \sum_{X,r,s} d_{XL,\beta}^{rs}  [\epsilon_X^{rs}]_{\beta\gamma} \delta_{\beta\delta}
          + \sum_{X,r,s} d_{XL,\beta}^{rs*} \delta_{\beta\gamma} [\epsilon_X^{rs}]^*_{\beta\delta}
          + \sum_{X,Y,r,s} d_{XY,\beta}^{rs} [\epsilon_X^{rs}]_{\beta\gamma}
                                             [\epsilon_Y^{rs}]^*_{\beta\delta} \bigg] \,.
  \label{eq:tildeP-L0}
\end{align}}%
We see that $\tilde P^S_{\alpha\beta} (E_\nu)_{L=0}$ depends only on the EFT parameters $\epsilon_X$ weighted by the appropriate production and detection coefficients, while the the dependence on $U_{\alpha n}$ has been eliminated by using unitarity of the PMNS matrix.

%-----------------------------------------------------------------------
\subsection{Neutrino Production in Meson Decays}
\label{sec:production}
%-----------------------------------------------------------------------

Essentially all neutrinos observed in FASER$\nu$ are produced in decays of light mesons and baryons.  
Details on FASER$\nu$'s neutrino fluxes are given in \cref{sec:FASER} and in ref.~\cite{Abreu:2019yak}, see in particular Table~1 in that reference. 
It was found that the dominant flux
reaching the FASER$\nu$ detector consists mainly of (anti-)neutrinos from pion and kaon decays. 
In the following we derive the production coefficients for these two dominant sources.
We also derive the production coefficients for $D_s$ meson decays, which are the main source of $\tau$ neutrinos in FASER$\nu$. 
Our discussion concerns neutrino production, but for anti-neutrinos the analytic results are the same up to negligible effects due to CP violation in the neutral kaon system.

\subsubsection{Pion Decay}
\label{sec:pion-decays}
%-----------------------------

Charged pion decays are mediated at the parton level by $u  \bar d \to \ell^+_\alpha \nu$, thus, they are sensitive to the $[\epsilon_X^{ud}]_{\alpha\beta}$ Wilson coefficients in the WEFT Lagrangian of \eqref{eq:EFT_lweft}.  
In our formalism, this translates to pion decays contributing to the $p_{XY,\alpha}^{\pi,ud}$ production coefficients, whereas $p_{XY,\alpha}^{\pi,jk} = 0$ for $jk \neq ud$. 
The main decay channel is 2-body: $\pi^+ \to \ell^+_\alpha \nu$, while other channels have tiny branching fractions and can be safely neglected here. 
This leads to a tremendous simplification because 2-body matrix elements depend only on the masses of the involved particles and not on the kinematics. 
Pulling the amplitudes in front of the integrals in \eqref{eq:pxy}, the integrals cancel between the numerator and the denominator, and we are left with a compact expression:  
\begin{align}
    p_{XY,\alpha}^{\pi,ud} =
        \frac{A_{X,\alpha}^{\pi,ud} A_{Y,\alpha}^{\pi,ud*}}
             {|A_{L,\alpha}^{\pi,ud}|^2} \,.
    \label{eq:pxy-faser2b}
\end{align}
Thanks to this simplification, the coefficients $p_{XY,\alpha}^{\pi,ud}$ only depend on the fundamental physics encapsulated by the amplitudes $A_{X,\alpha}^{\pi,ud}$. In particular they are independent of the energy distribution of the parent pions.  

To evaluate $p_{XY,\alpha}^{\pi,ud}$, we need the pion decay amplitude, both in the SM and in the presence of new physics with non-standard Lorentz structures.  
Regarding the latter, we can infer from the quantum numbers of the charged pion, $J^P = 0^-$, that only the axial-vector current $\bar{u} \gamma^\mu \gamma^5 d$ and the pseudoscalar current $\bar{u} \gamma^5 d$ can have non-zero matrix elements. The vector and  scalar currents  are parity-even; for the tensor current, one can  argue that no antisymmetric tensor can be formed from the only available Lorentz vector in the problem, namely $p^\mu_\pi$, the pion 4-momentum.
All in all, the amplitudes entering in \eqref{eq:pxy-faser2b} can be expressed as 
\begin{align}
  \label{eq:PION_AP1}
    A^{\pi,ud}_{L,\alpha} &= -A^{\pi, ud}_{R,\alpha}
        = \frac{V_{ud}}{v^2} (\bar{u}_{\nu} \gamma^\mu P_L v_{\ell_\alpha})
          \bra{0} \bar{d} \gamma_\mu \gamma_5 u \ket{\pi^+(p_\pi)} \,,  \nnl
    A^{\pi,ud}_{P,\alpha} &=
         -\frac{V_{ud}}{v^2} (\bar{u}_{\nu} P_R v_{\ell_\alpha})
          \bra{0} \bar{d} \gamma_5 u \ket{\pi^+(p_\pi)} \,,  \nnl 
    A^{\pi,ud}_{S,\alpha} &= A^{\pi,ud}_{T,\alpha} = 0 , 
\end{align}
where $v_{\ell_\alpha}$, $\bar{u}_{\nu}$ are the Dirac spinor wave functions of the charged lepton and the neutrino, respectively.\footnote{Note that $u$ in $u_{\nu} $ stands for the positive energy solution of the Dirac equation and should not be confused with the field operator $u$ of the up-quark field.}
The hadronic matrix elements in \eqref{eq:PION_AP1} are customarily parameterized as~\cite{Aoki:2019cca}\footnote{%
The second expression in \eqref{eq:pi-decay-matrix-elements} can be obtained from the first one by contracting the latter with $p_\pi^\mu$ and using the equations of motion. 
} 
\begin{align}
    \bra{0} \bar{d} \gamma^\mu \gamma_5 u \ket{\pi^+(p_\pi)}
        = i p_\pi^\mu f_{\pi} \,,
    \qquad 
    \bra{0} \bar{d} \gamma_5 u \ket{\pi^+(p_\pi)}
        = -i \frac{m_{\pi}^2}{m_u + m_d}  f_{\pi} \,.
  \label{eq:pi-decay-matrix-elements}
\end{align}
where $f_{\pi} = 130.2(0.8)$~MeV~\cite{Aoki:2019cca} is the pion decay constant (which will cancel from our final results), $m_{\pi}$ is the charged pion mass, and $m_u$, $m_d$ are the up and down quark masses.  
With \cref{eq:pi-decay-matrix-elements}, the left-handed and pseudoscalar amplitudes  become 
\begin{align}
    A_{L,\alpha}^{\pi,ud}
        = -i \frac{V_{ud}}{v^2} (\bar{u}_{\nu} P_R v_{\ell_\alpha})
          (f_\pi m_{\ell_\alpha}) \,,
    \qquad
    A_{P,\alpha}^{\pi,ud}
        = i \frac{V_{ud}}{v^2} (\bar{u}_{\nu} P_R v_{\ell_\alpha})
          \frac{f_\pi m_\pi^2}{m_u + m_d} \,, 
\end{align} 
and their squares, summed over spins, are
\begin{align}
\label{eq:PION_sumAXY}
    \sum |A_{L,\alpha}^{\pi,ud}|^2
        &= \frac{V_{ud}^2 f_{\pi}^2}{v^4} m_{\ell_\alpha}^2
           \big( m_{\pi}^2 - m_{\ell_\alpha}^2 \big) \,, \nnl 
    \sum A_{L,\alpha}^{\pi,ud} \bar A_{P,\alpha}^{\pi,ud}
        &= -\frac{V_{ud}^2 f_{\pi}^2}{v^4} m_{\ell_\alpha}
            \big( m_{\pi}^2 - m_{\ell_\alpha}^2 \big)
            \frac{m_{\pi}^2}{m_u + m_d} \,, \nnl 
    \sum |A_{P,\alpha}^{\pi,ud}|^2
        &= \frac{V_{ud}^2 f_{\pi}^2}{v^4}
           \big(  m_{\pi}^2 - m_{\ell_\alpha}^2 \big)  
           \frac{m_{\pi}^4}{(m_u + m_d)^2} \,.
\end{align} 
Plugging \cref{eq:PION_sumAXY} into \cref{eq:pxy-faser2b}, we obtain the production coefficients~\cite{Falkowski:2019kfn}
\begin{align}
p_{LL,\alpha}^{\pi,ud} &=  p_{RR,\alpha}^{\pi,ud} = -p_{LR,\alpha}^{\pi,ud} = 1
                                                        \,,\notag\\[0.2cm]
p_{PL,\alpha}^{\pi,ud} &= -p_{PR,\alpha}^{\pi,ud}
                   = -\frac{m_\pi^2}{m_{\ell_\alpha} (m_u+m_d)} 
                   \simeq -27\ (-5600)
                          &\quad\text{for $\alpha=\mu\ (e)$} \,,
                                                     \label{eq:pioncoefficients} \\
p_{PP,\alpha}^{\pi,ud} &= \frac{m_\pi^4}{m_{\ell_\alpha}^2 (m_u+m_d)^2}
                   \simeq 730 \ (3.1 \times 10^7)
                          &\quad\text{for $\alpha=\mu\ (e)$} \,, \notag
\end{align}
where in the numerical evaluation we have used 
$m_u+m_d = \SI{6.82(9)}{MeV}$~\cite{Aoki:2019cca}. 
The factor $m_\pi/m_{\ell_\alpha}$ appears in the production coefficients involving pseudoscalar interactions because they, unlike the SM ones, do not suffer from chiral suppression. 
Thus, even a small pseudoscalar coupling $[\epsilon_P^{ud}]_{\alpha \gamma}$ (with $\alpha = \mu, e$) present in the effective Lagrangian \eqref{eq:EFT_lweft} leads to a significant enhancement of the FASER$\nu$ neutrino fluxes compared to the SM prediction. 
This enhancement will allow us to obtain particularly strong constraints ($\sim 10^{-3}$) on some of the Wilson coefficients $\epsilon_P^{ud}$.
Note, however, that in many extension of the SM the Wilson coefficients corresponding to new scalar and pseudoscalar operators are expected to scale as the corresponding quark masses, which would cancel this enhancement
when considering the sensitivity to the parameters of the ultraviolet-complete BSM model.
We also note that the bare quark masses appearing in \cref{eq:pioncoefficients} are heavily dependent on the renormalization scale. 
However, this scale dependence is canceled 
by the scale dependence of the Wilson coefficients, allowing us to set robust constraints in spite of the scale dependence.

\subsubsection{Kaon Decay}
\label{sec:kaon-decays}
%--------------------------

The second most important contribution to FASER$\nu$'s neutrino flux comes from kaon decays.
At the parton level the relevant decays are mediated by $u \bar s \to \ell^+_\alpha \nu$ transitions,
therefore they are sensitive to the Wilson coefficients $[\epsilon_X^{us}]_{\alpha \beta}$.  
The evaluation of the corresponding production coefficients $p_{XY,\alpha}^{K,us}$ is more involved than in the pion case.
The reason is that relevant decay channels are both 2-body ($K^+\to \ell_\alpha^+ \nu$) and 3-body ($K^+\to \pi_0 \ell_\alpha^+ \nu$, $K_{L,S} \to \pi^- \ell_\alpha^+ \nu$). 
Moreover, the kinematics and phase space integrations are quite cumbersome for 3-body decays.   
In the following we only quote the results, leaving more details of the derivation for \cref{app:kaon}. 

We write the production coefficients in the form 
\begin{align}
    p_{XY,\alpha}^{K,us} 
=   \frac{N_{XY}}{D_{LL}^\text{2b} + D_{LL}^\text{3b}}\,, 
    \label{eq:PROD_pusXYmu_ND}
\end{align}
where $N_{XY}$ stands for the numerator in \eqref{eq:pxy}, and $D_{LL}^\text{2,3b}$ stand for 2-body and 3-body contributions to the denominator in \eqref{eq:pxy}.

Starting with the denominator, the 2-body contribution due to the $K^+ \to \ell^+_\alpha \nu$ decay is given by  
\begin{align}
    D_{LL}^\text{2b}
        = \frac{V_{us}^4 f_{K}^2 }{8\pi v^4} m_{\ell_\alpha}^2
          \big( m_{K^\pm}^2 - m_{\ell_\alpha}^2 \big)
          \int\!dE_K \frac{\phi_{K_+}(E_K)}{p_K E_K}
              \Big[ \Theta\big( E_\nu - E_{\nu}^\text{min} \big) 
                  - \Theta\big( E_\nu - E_{\nu}^\text{max} \big)
              \Big] \,,
    \label{eq:KAON_DLL2b}
\end{align}
where $p_K = \sqrt{E_K^2 - m_K^2}$,  $\phi_{K^+}(E_K)$ is the energy distribution of the parent $K^+$ mesons, $\Theta(x)$ is the Heaviside step function, the charged kaon decay constant is $f_{K} = 155.7(3)$\,MeV~\cite{Aoki:2019cca}, and 
\begin{align}
    E_{\nu}^\text{min} 
        = \frac{m_{K} ^2 - m_{\ell_\alpha}^2}{2 (E_K - p_K \cos\theta_0 )}\,, 
    \qquad 
    E_{\nu}^\text{max}
        = \frac{m_{K}^2 - m_{\ell_\alpha}^2}{2 (E_K - p_K)} \,.
\end{align} 
Above, $\theta_0$ is the kinematic cut on the direction of the emitted neutrino in the lab frame, relative to the beam axis. In our numerical analysis we use the value $\theta_0 = 5.6 \times 10^{-4}$, based on the geometry of the FASER$\nu$ detector~\cite{Abreu:2019yak}. The factor $1/p_K$ in \cref{eq:KAON_DLL2b} comes from the 2-body phase space integral, which can be written as $d\Pi_\text{2b} = d\cos\theta_{\nu,\text{cm}} E_{\nu,\text{cm}} / (8 \pi m_{K}) = dE_\nu / (8 \pi p_K)$, where $\theta_{\nu,\text{cm}}$ is the angle between the direction into which the neutrino is emitted in the center-of-mass frame and the beam axis, and $E_{\nu,\text{cm}}$ is its energy in that frame, while $E_\nu$ is its energy in the lab frame. 

The 3-body contribution  to the denominator arises due to semileptonic kaon decays.
We find 
\begin{align} 
    D_{LL}^\text{3b} = \frac{1}{128 \pi^3}
                       \int_{E_K^\text{min}}^{\infty} \frac{dE_K}{E_K}
                       \sum_i \phi_{K_i}(E_K) \int_{\cos\theta_{\min} }^{1}
                       \frac{d\cos\theta}{E_K - p_K \cos\theta}
                       \int_{q^2_{\min}}^{q^2_{\max}} \! dq^2
                       \sum_{\rm spin} |A_{L,\alpha}^{K_i \, us} |^2 , 
 \label{eq:KAON_DLL3b}
\end{align}
where $\sum_i$ runs over semileptonic $K_+$, $K_S$, $K_L$ decays, and 
\begin{align}
    \cos\theta_{\min} &=
        \max \bigg[ \cos\theta_0,\, \frac{2 E_K E_\nu - m_K^2 + (m_\pi + m_{\ell_\alpha})^2}
                                         {2 p_K E_\nu}  \bigg ] \,,  
\nnl 
    E_K^\text{min} &=
        \frac{m_K^2}{m_K^2 - (m_\pi + m_{\ell_\alpha})^2}  E_\nu \,, 
  \qquad  w^2 =
        m_K^2 + 2 E_\nu (p_K \cos\theta - E_K) \,.  
\end{align}
The $q^2$ integration limits are given in \cref{eq:KAON_qsqminmax}.  
The amplitude squared occurring in \cref{eq:KAON_DLL3b} is given in \cref{eq:KAON_sumAXY-semileptonic}. 

For the numerators in \cref{eq:PROD_pusXYmu_ND} we have 
\begin{align}
    N_{LL}  &= N_{RR} = D_{LL}^\text{2b} + D_{LL}^\text{3b} \,,
    \notag\\[0.2cm] 
    N_{LR}  &= -D_{LL}^\text{2b} + D_{LL}^\text{3b} \,, 
    \notag\\[0.1cm]
    N_{LP}  &= -N_{RP} = -\frac{m_{K^\pm}^2}{m_{\ell_\alpha} (m_u + m_s)}  D_{LL}^\text{2b} \,,
    \nnl 
    N_{PP}  &= \frac{m_{K^\pm}^4}{m_{\ell_\alpha}^2 (m_u + m_s)^2} D_{LL}^\text{\rm 2b} ,
    \label{eq:KAON_NXY2b}
\end{align} 
and  
\begin{align}
    N_{XY} = \frac{1}{128 \pi^3}  \int_{E_K^\text{min}(E_\nu)}^{\infty}
             \frac{dE_K}{E_K} \sum_i \phi_i(E_K)
             \int_{\cos\theta_{\min}}^{1}
             \frac{d\cos\theta}{E_K -  p_K \cos\theta}
             \int_{q^2_{\min}}^{q^2_{\max}} dq^2
             \sum_\text{spin} A_{X,\alpha}^{K_i,us}  \bar{A}_{Y,\alpha}^{K_i,us} ,
    \label{eq:KAON_NXY3b}
\end{align}
for $XY = LS, LT,SS,TT,ST$. 
The expressions for the amplitudes needed in \cref{eq:KAON_NXY3b} are collected in \cref{eq:KAON_sumAXY-semileptonic}.  
Moreover, $N_{RS} = N_{LS}$ and $N_{RT} = N_{LT}$, $N_{SP} = N_{TP} = 0$.

\begin{figure}
  \centering
  \includegraphics[width=1\textwidth]{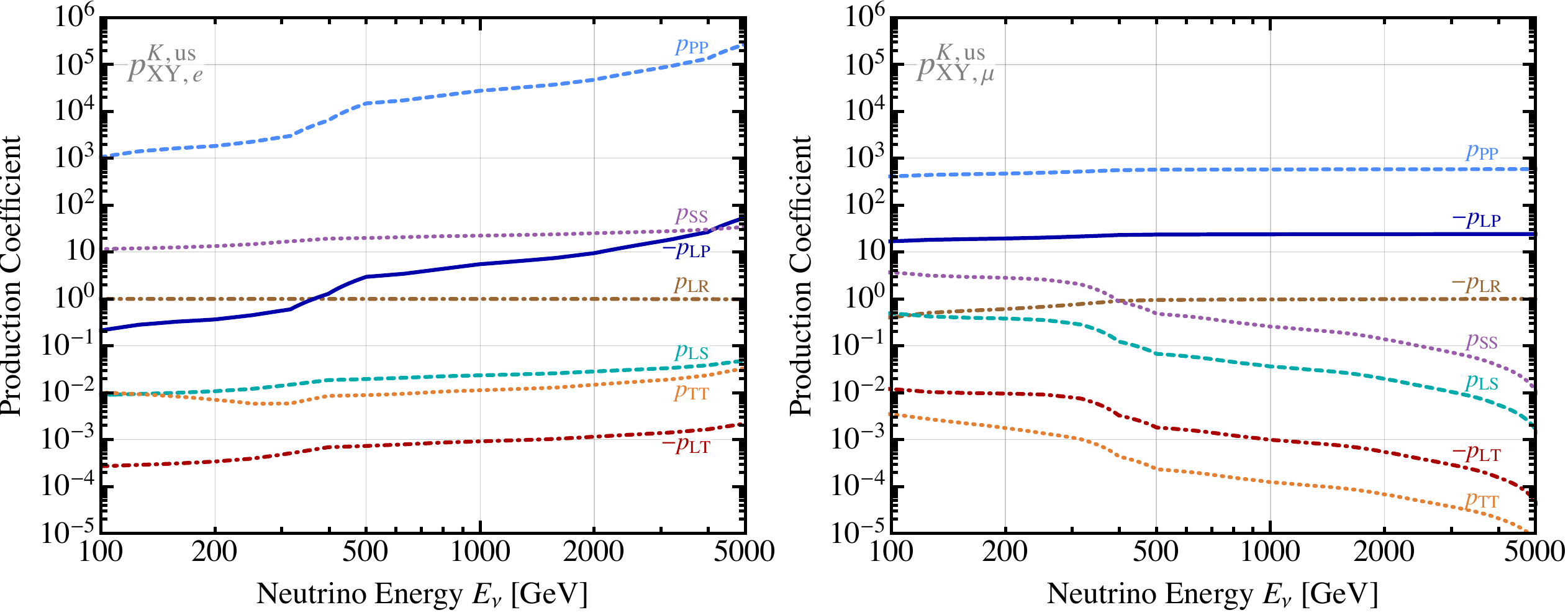}
  \caption{Production Coefficients for kaon decay. Left: decay to electrons. Right: decay to muons.}
  \label{fig:pXY}
\end{figure}

The production coefficients relevant to our work are plotted in \cref{fig:pXY} as a function of neutrino energy.
For muonic decays of kaons the flux at the relevant neutrino energies is dominated by the leptonic channel. 
For this reason $p_{LR,\mu}^{K,us}$, $p_{LP,\mu}^{K,us}$, and $p_{PP,\mu}^{K,us}$ are almost flat in energy, similar to the production coefficients for purely leptonic decays of pions, cf.\ \cref{eq:pioncoefficients}. 
The effect of the semileptonic  admixture is to produce non-zero production coefficients associated with the scalar and tensor interactions:  $p_{LS,\mu}^{K,us}$, $p_{LT,\mu}^{K,us}$,   $p_{SS,\mu}^{K,us}$, etc. 
Their shape, quickly decreasing with $E_\nu$, is due to the fact that the domination of leptonic over semileptonic fluxes becomes stronger at high energy. 
Conversely, for electronic decays of kaons,  semileptonic channels largely dominate over the chirally suppressed $K \to e \nu$. 
Therefore the production coefficients associated with the scalar and tensor interactions are approximately constant, their variation being due to the energy dependence of the corresponding matrix elements. 
The production coefficients associated with the pseudoscalar interactions,  $p_{LP,e}^{K,us}$, and $p_{PP,e}^{K,us}$ are still relatively large, because they are chirally enhanced by $m_K/m_e$. 
Their sharp increase with energy is due to the semileptonic fluxes quickly shutting off at large $E_\nu$, which leads to the relative contribution of the leptonic decay becoming more important at higher energies.

%--------------------------
\subsubsection{Charm Decay}
\label{sec:charm-decays}

The decays $D_s \to \ell^+_\alpha \nu$ are mediated at the parton level by $c  \bar s \to \ell^+_\alpha \nu$, thus they are  sensitive to the $[\epsilon_X^{cs}]_{\alpha \beta}$ Wilson coefficients in the EFT Lagrangian of \eqref{eq:EFT_lweft}.
While their contribution to the neutrino flux is much smaller than that of pion and kaon decays, they are the main source of tau neutrinos in FASER$\nu$. 
Consequently, they give us access to the $[\epsilon_X^{cs}]_{\tau \gamma}$ Wilson coefficients, which cannot be probed by pion and kaon decays.   
The calculation is completely analogous to the one for pion decay in Sec.~\ref{sec:pion-decays}.  
We find the non-vanishing $p_{XY,\alpha}^{D,cs}$ production coefficients: 
\begin{align}
p_{LL,\alpha}^{D,cs} &=  p_{RR,\alpha}^{D,cs} = -p_{LR,\alpha}^{D,cs} = 1
                                                        \,,\notag\\[0.2cm]
p_{PL,\alpha}^{D,cs} &= -p_{PR,\alpha}^{D,cs}
                   = -\frac{m_{D_s}^2}{m_{\ell_\alpha} (m_c+m_s)} 
                   \simeq -1.6,\,-27,\,-5.5 \times 10^3 
                          &\quad\text{for $\alpha=\tau,\,\mu,\,e$} \,,
                                                     \label{eq:Dscoefficients} \\
p_{PP,\alpha}^{D,cs} &= \frac{m_{D_s}^4}{m_{\ell_\alpha}^2 (m_c+m_s)^2}
                   \simeq  2.5,\,710,\,3.0 \times 10^7 
                          &\quad\text{for $\alpha=\tau,\,\mu,\,e$} \,. \notag
\end{align}
In the numerical evaluation we used $m_c = 1.280(13)$~GeV and $m_s = 92.9(7)$~MeV~\cite{Zyla:2020zbs}.

We will neglect new physics contributions to neutrino production in $D^\pm$ and $D^0$ decays because they only make very small contributions to the overall neutrino flux.

%-----------------------------------------------------------------------
\subsection{Neutrino Detection via Deep-Inelastic Scattering}
\label{sec:detection}
%-----------------------------------------------------------------------

At FASER$\nu$ energies, neutrino detection proceeds almost exclusively
through charged-current deep-inelastic scattering~(DIS),
\begin{align}
  \nu + N \to \ell + X \,,
\end{align}
where $N = n, p$ is a nucleon and $X$ can be any hadronic final state.

\subsubsection{Deep-Inelastic Scattering in the Standard Model}
%--------------------------------------------------------------

In the SM, the differential cross section for deep-inelastic charged-current neutrino scattering on a nucleon $N$ is
\begin{align}
  \frac{d^2\sigma^{\text{SM}}_{\beta N}}{dx\,dQ^2}
    &= \frac{1}{2 \pi v^4}  \bigg[ 
    \bigg(1-\frac{m_{\ell_\beta}^2}{\hat s}\bigg) \sum_{q=d,s}  f^N_q(x,Q^2)
                     + \frac{(\hat s - Q^2) (\hat s - Q^2-m^2_{\ell_\beta})}
                            {\hat s^2}
                       \sum_{\bar{q}=\bar{u},\bar{c}} f^N_{\bar{q}}(x,Q^2) \bigg]
  \label{eq:SMxsection-nu} \\
\intertext{for neutrino scattering, and}
  \frac{d^2\bar \sigma^{\text{SM}}_{\beta N}}{dx\,dQ^2}
    &= \frac{1}{2 \pi v^4}  \bigg[  
\bigg(1-\frac{m_{\ell_\beta}^2}{\hat s }\bigg) \sum_{\bar q=\bar d,\bar s} f^N_q(x,Q^2)
+ \frac{(\hat s - Q^2)(\hat s - Q^2 - m^2_{\ell_\beta})}{\hat s^2}
\sum_{q=u,c} f^N_{q}(x,Q^2) \bigg]
  \label{eq:SMxsection-nubar}
\end{align}
for anti-neutrino scattering. 
In these expressions, we have set all quark masses to zero. 
$m_{\ell_\beta}$ denotes the mass of the outgoing charged lepton of flavor $\beta$ (which is not completely negligible for $\beta = \tau$) and
\begin{align}
\hat s = \frac{4 m_N E_\nu^2}{2 E_\nu +  m_N} x \approx 2  m_N E_\nu x . 
  \label{eq:E-nu-prime}
\end{align}
is the invariant mass squared of the neutrino--quark system. 
Next, $x$ is  the fraction of the nucleon momentum carried by the incident quark, and $Q^2 \equiv - (p_\nu - p_{\ell_\beta})^2$ is the invariant momentum transfer. 
The functions $f^N_q(x, Q^2)$ are the PDFs of the nucleons.  

\begin{figure}
  \centering
  \includegraphics[width=0.8\textwidth]{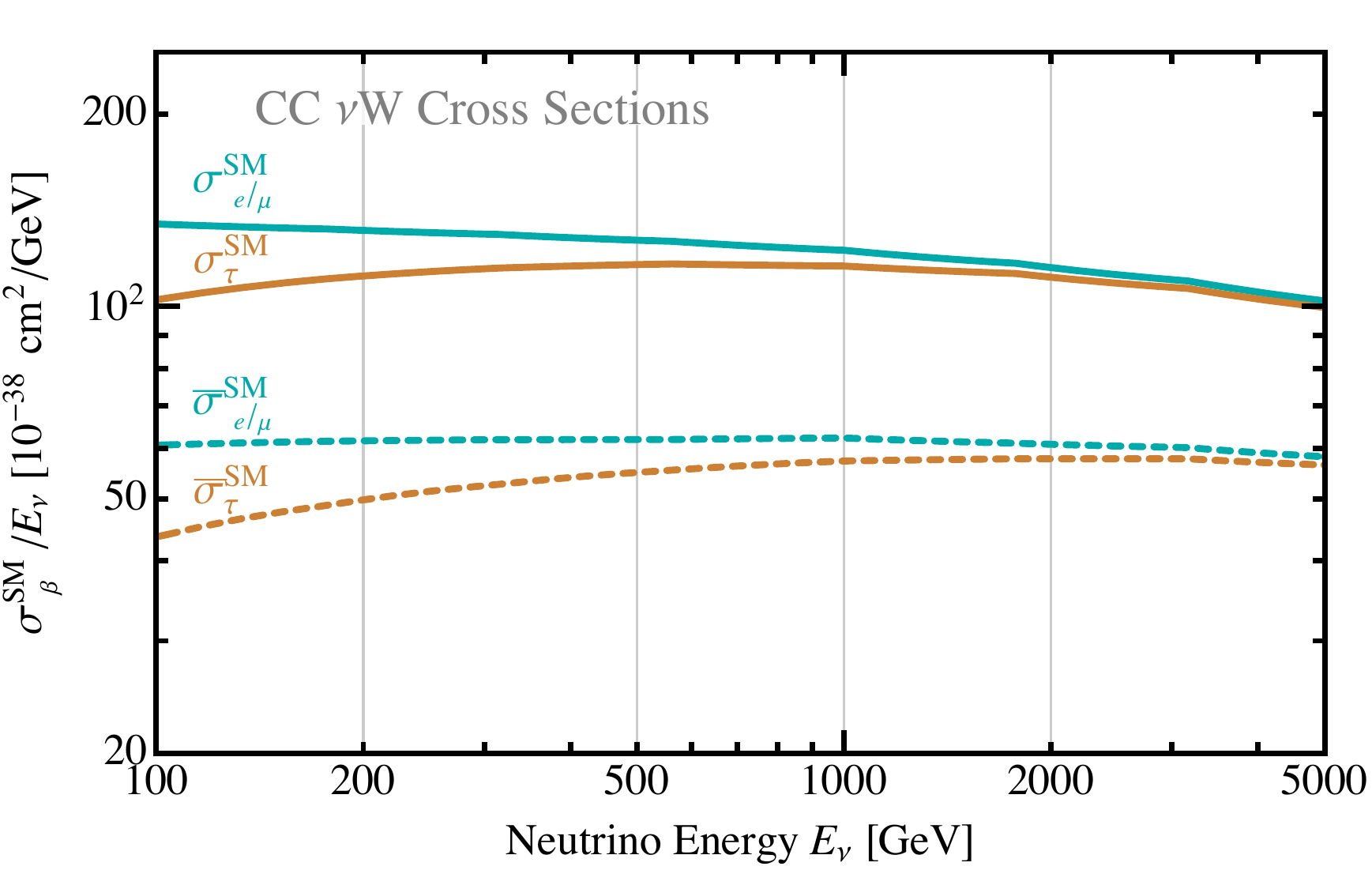}
  \caption{% 
  SM cross sections for neutrino (solid) and anti-neutrino (dashed) charged-current (CC) deep-inelastic scattering on tungsten. 
  We have used the LO~MSTW~PDFs \cite{Martin:2009iq}, see text for details. 
  }
  \label{fig:SMxsection}
\end{figure}

The SM limit of the detection cross section for a neutrino scattering on a nucleus, which is denoted as $\sigma^{\text{SM}}_\beta(E_\nu)$ in  the master formula of \cref{eq:masterRate}, 
is obtained by integrating \eqref{eq:SMxsection-nu}  over the $x$ and $Q^2$ variables and summing over the nucleons:  
\begin{align}
    \sigma^{\text{SM}}_\beta(E_\nu) 
=   \sum_{N=n,p} n_N \int_{x_0}^1 d x 
    \int_{Q^2_0}^{\hat s - m_{\ell_\beta}^2 }  d Q^2 
    \frac{d^2\sigma^{\text{SM}}_{\beta N}}{dx\,dQ^2}, 
\end{align}
where $Q_0 \sim \SI{1}{GeV}$ and $x_0 = (Q_0^2 + m_{\ell_\beta}^2)/(2  m_N E_\nu)$. 
An analogous formula holds for the anti-neutrino cross section $\bar \sigma^{\text{SM}}_\beta(E_\nu)$. 
The target nucleus in FASER$\nu$ is tungsten with  $n_p = 74$ protons and on average $n_n \simeq 110$ neutrons. 
The results of the numerical integration  are shown in \cref{fig:SMxsection} as a function of the incident neutrino energy. 
For the proton PDFs we used the MSTW set~\cite{Martin:2009iq} at leading order.
The neutron PDFs are related to the proton ones by the isospin symmetry: 
$f^n_u = f^p_d$ and $f^n_d = f^p_u$. 
The distributions of the strange and charm quarks and anti-quarks are the same for protons and neutrons.
We find a good  agreement with ref.~\cite{Abreu:2019yak}.

\subsubsection{Deep-Inelastic Scattering in EFT Extensions of the Standard Model}
%------------------------------------------------------------------

We move to discussing the effects of new physics on the detection side, encoded in  the detection coefficients defined in \cref{eq:dxy}. 
To calculate these, we need the amplitudes for a neutrino scattering on a quark inside a target nucleon in the presence of non-SM interaction in \cref{eq:EFT_lweft}. 
For $\nu_n d^k \to \ell_\beta^- u^j$ the amplitude decomposes as in \cref{eq:Mdecomposition} with the reduced amplitudes given by 
\begin{align}
    \begin{split}
        A_{L,\beta}^{D,jk} &=
            - \frac{2 V_{jk}}{v^2} (\bar u_u \gamma^\mu P_L u_d)(\bar u_{\ell_\beta} \gamma_\mu P_L u_\nu) \,, \\
        A_{R,\beta}^{D,jk} &=
            - \frac{2 V_{jk}}{v^2} (\bar u_u \gamma^\mu P_L u_d)(\bar u_{\ell_\beta} \gamma_\mu P_L u_\nu) \,, \\ 
        A_{S,\beta}^{D,jk} &= -\frac{V_{jk}}{v^2} (\bar u_u u_d)(\bar u_{\ell_\beta} P_L u_\nu) \,, \\ 
        A_{S,\beta}^{D,jk} &= \frac{V_{jk}}{v^2} (\bar u_u \gamma_5 u_d)(\bar u_{\ell_\beta} P_L u_\nu) \,, \\
        A_{S,\beta}^{D,jk} &=
            - \frac{V_{jk}}{2 v^2} (\bar u_u \sigma^{\mu\nu} u_d)(\bar u_{\ell_\beta} \sigma_{\mu\nu} P_L u_\nu) \,, 
    \end{split}
    \label{eq:ADjk}
\end{align}
where $u_i$ are the spinor wave functions of the involved quarks and leptons. 
For $\nu_n \bar u^j \to \ell_\beta^- \bar d^k$ the reduced amplitudes $A_{L,\beta}^{\bar D,jk}$ are obtained from the above ones by replacing $u_q \to v_q$ (up to an irrelevant minus sign from Fermi statistics). 
For anti-neutrino scattering one needs to replace $u \to v$ in the lepton's wave functions and take the complex conjugate.   

In the limit where quarks are treated as massless, the spin-summed amplitudes squared are given by the compact expressions 
\begin{align}
    \begin{split}
        \sum_\text{spin} A_{L,\beta}^{D,jk} A_{L,\beta}^{D,jk*} &=
            \frac{16|V_{jk}|^2}{v^4} \hat s (\hat s  - m_{\ell_\beta}^2 ) \,, \\
        \sum_\text{spin} A_{L,\beta}^{D,jk} A_{X,\beta}^{D,jk*} &= 0, \hspace{4cm} (X = R,S,P,T) , \\
        \sum_\text{spin} A_{R,\beta}^{D,jk} A_{R,\beta}^{D,jk*} &=
            \frac{16|V_{jk}|^2}{v^4} \big (\hat s - Q^2 \big) \big(\hat s - Q^2 - m_{\ell_\beta}^2 \big) \,, \\ 
        \sum_{\rm spin} A_{S,\beta}^{D,jk} A_{S,\beta}^{D,jk*} &=
        \sum_{\rm spin} A_{P,\beta}^{D,jk} A_{P,\beta}^{D,jk*} = 
            \frac{2 |V_{jk}|^2}{v^4} Q^2  \big ( Q^2 + m_{\ell_\beta}^2  \big) \,, \\
        \sum_{\rm spin} A_{T,\beta}^{D,jk} A_{T,\beta}^{D,jk*} &=
            \frac{4|V_{jk}|^2}{v^4} \bigg[  (2 \hat s - Q^2)^2 -  m_{\ell_\beta}^2  (4 \hat s - Q^2) \bigg] \,. 
    \end{split}
    \label{eq:axy}
\end{align}
Note that, because we have taken the quark masses to be zero, all 
$\sum A_{L,\beta}^{D,jk} A_{X,\beta}^{D,jk*}$ with $X \neq L$, and thus all detection coefficients $d_{LX,\beta}^{jk}$ with $X \neq L$, vanish. 
This implies that the only Wilson coefficients that can modify the detection process at linear order are the $\epsilon_L^{jk}$. For neutrino--anti-quark scattering, we have
\begin{align}
    \sum_\text{spin}  A_{L,\beta}^{\bar D,jk} A_{L,\beta}^{\bar D,jk*}
    &= \sum_{\rm spin} A_{R,\beta}^{D,jk} A_{R,\beta}^{D,jk*} 
    \notag\\
 \sum_\text{spin}     A_{R,\beta}^{\bar D,jk} A_{R,\beta}^{\bar D,jk*}
    &=   \sum_\text{spin}  A_{L,\beta}^{D,jk} A_{L,\beta}^{D,jk*} \, , 
\end{align}
and the remaining $ \sum_\text{spin} A_{X,\beta}^{\bar D,jk} A_{Y,\beta}^{\bar D,jk*}$ are identical to their counterparts for neutrino--quark scattering.

\begin{table}
    \centering
    \begin{tabular}{cccc|ccc}
        \toprule
            & \multicolumn{3}{c|}{neutrinos} & \multicolumn{3}{c}{anti-neutrinos} \\
            \cmidrule{2-4} \cmidrule{5-7}
            & $d_{XX,\beta}^{ud}$      & $d_{XX,\beta}^{us}$     & $d_{XX,\beta}^{cs}$ 
            & $\bar d_{XX,\beta}^{ud}$ & $\bar d_{XX,\beta}^{us}$  & $\bar d_{XX,\beta}^{cs}$  \\
        \midrule
         L   & 0.91 & $5.2\times10^{-3}$    & $6.9\times10^{-2}$  & 0.82 & $3.1\times10^{-2}$     & 0.14 \\
         R   & 0.45 & $8.6\times10^{-3}$   &$7.2\times10^{-2}$  & 1.61 &  $7.8\times10^{-2}$   & 0.15 \\
         S/P & 0.04 & $4.5\times10^{-4}$  & $5.5\times10^{-3}$ & 0.07 & $3.3\times10^{-3}$   & 0.01 \\
         T   & 0.59 & $6.0\times10^{-3}$    & $6.7\times10^{-2}$  & 1.07  & $4.7\times10^{-2}$    & 0.12 \\
        \bottomrule
    \end{tabular}
    \caption{Detection coefficients for the scattering of neutrinos and anti-neutrinos with energy $E_\nu = \SI{1}{TeV}$ on tungsten. The values shown are for $\beta = e$, but they are practically equal for $\beta = \mu,\tau$.}
    \label{tab:dxy}
\end{table} 

\begin{figure}
  \centering
  \includegraphics[width=\textwidth]{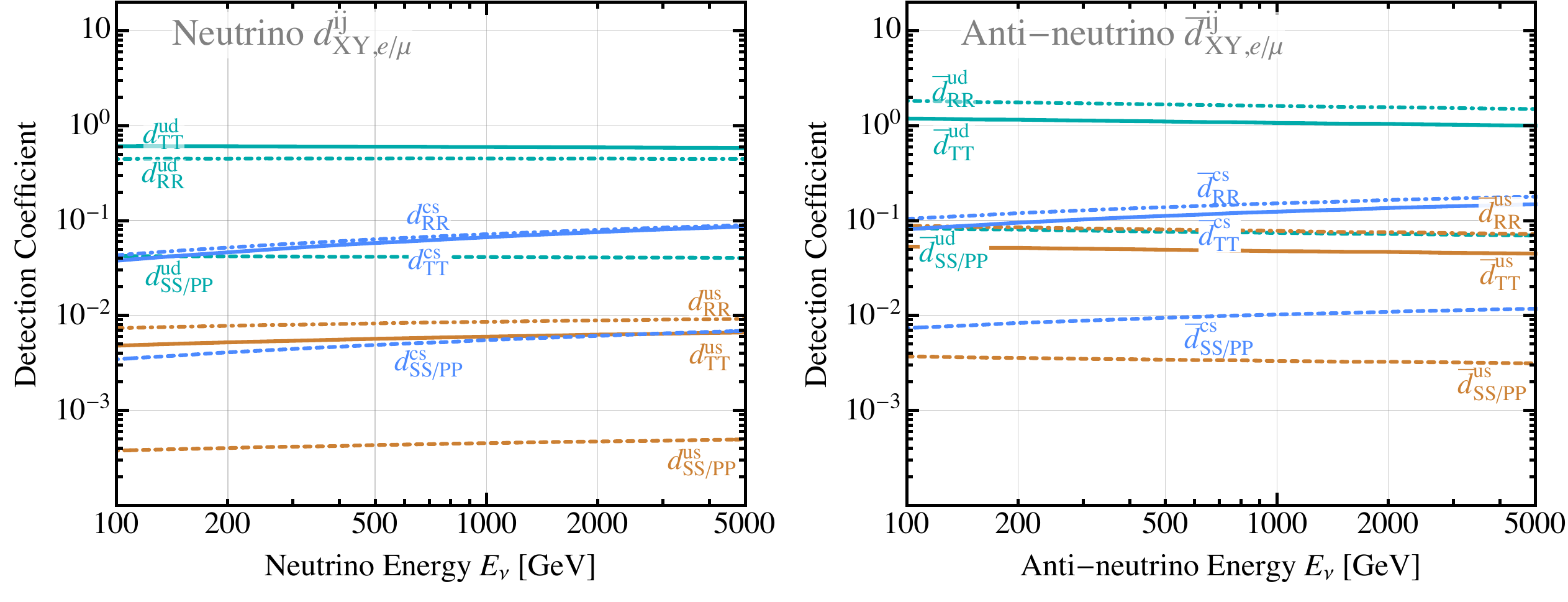}
  \caption{Energy dependence of the detection coefficients defined in \cref{eq:dxy}. Solid curves correspond to tensor couplings, dashed curves to scalar and pseudoscalar couplings, and dot-dashed curves to right-handed couplings.}
  \label{fig:dXY}
\end{figure}

To calculate the detection coefficients we can now insert \cref{eq:axy} into \cref{eq:dxy} and evaluate numerically the integral over the nucleon PDFs. 
The results for a particular neutrino energy $E_{\nu} = \SI{1}{TeV}$ are shown  in \cref{tab:dxy}.
For each Lorentz structure $X$, the largest  $d_{XX,\beta}^{jk}$
corresponds to the $jk = ud$ quark structure, which profits from the large PDFs of the up and down quarks in the nucleons and from the lack of Cabibbo suppression.
We note that, in most cases, the detection coefficients for anti-neutrinos are larger than the ones for neutrinos.
One reason is that the detection coefficients are inversely proportional to the SM scattering rate, which is roughly three times smaller for anti-neutrinos than for neutrinos. 
For right-handed couplings, the numerator in \cref{eq:dxy} is moreover larger for anti-neutrinos than for neutrinos. 
The scalar and pseudo-scalar coefficients are suppressed by small numerical factors, while the right-handed and tensor ones are much larger. 
This translates to a better sensitivity to the $\epsilon_R^{ud}$ and $\epsilon_T^{ud}$ Wilson coefficients on the detection side.
As for quark flavor structures other than $ud$, the detection coefficients $d_{XY,\beta}^{cs}$ are suppressed by the small PDFs of strange and charm quarks,\footnote{Note, however, that FASER$\nu$ has the capability of tagging charm mesons. This could be used to reduce the background to interactions of the form $\nu + s \to \ell^- + c$, recovering some sensitivity to $d_{XY,\beta}^{cs}$.} while the $d_{XY,\beta}^{us}$ are suppressed by the Cabibbo angle squared. 
Consequently, the sensitivity to $\epsilon_X^{us}$ and $\epsilon_X^{cs}$ is weak on the detection side.
The coefficients $d_{XY,\beta}^{cd}$ are suppressed by both the charm PDF and the Cabibbo angle, and we therefore do not consider them any further in this work.
The dependence of the detection coefficients on the incident neutrino energy is shown in \cref{fig:dXY}.
Most of the detection coefficients are, to a good approximation, energy-independent.  
A dependence on $E_\nu$ appears  due to the lepton masses, $Q^2$ dependence of the PDFs, and subleading terms in the relation between $\hat s$ and $E_\nu$, which are small effects at energies relevant for FASER$\nu$.

%=======================================================================
\section{Predicting the Sensitivity of FASER$\nu$}
\label{sec:FASER}
%=======================================================================

\begin{figure}
  \centering
  \includegraphics[width=\textwidth]{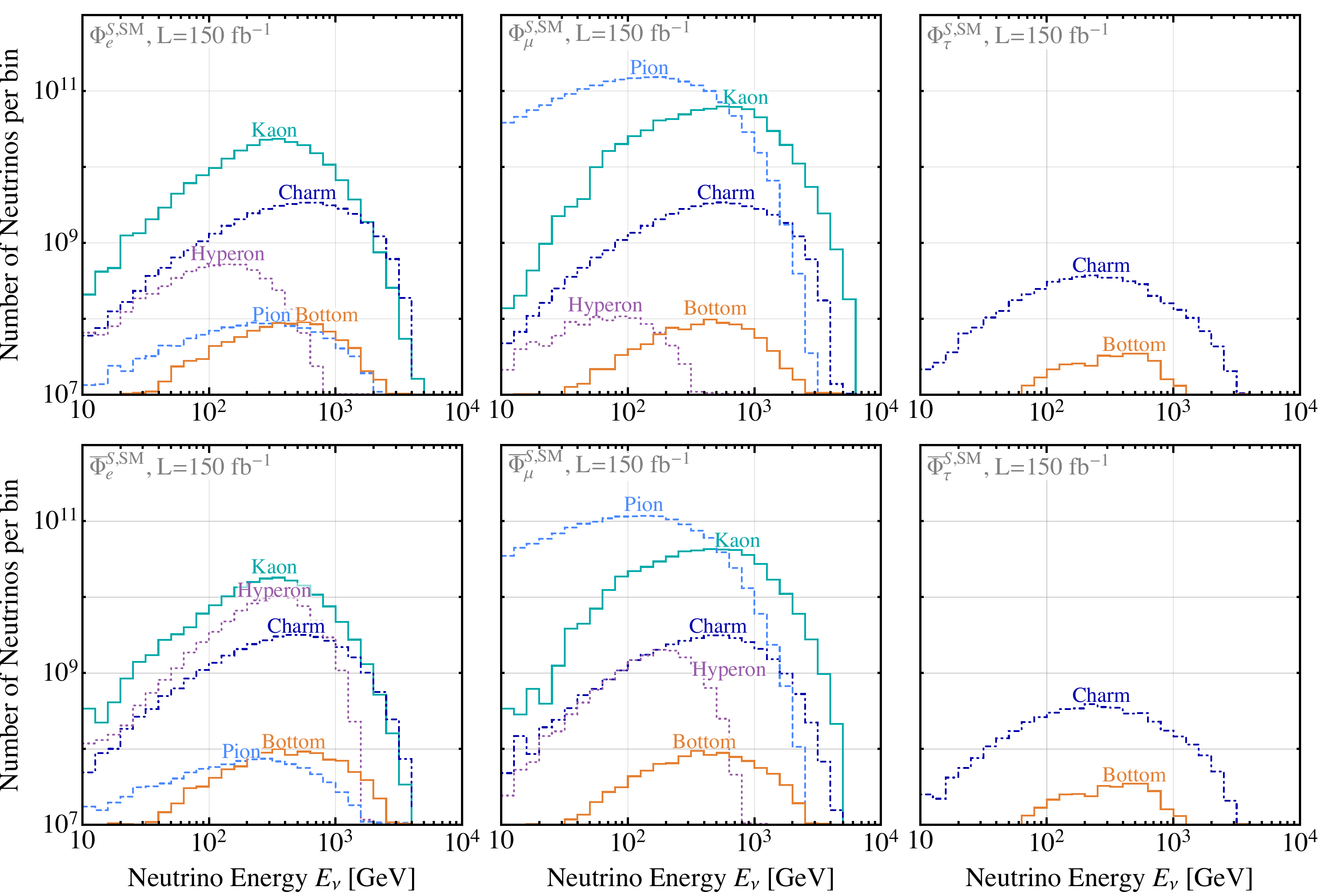}
  \caption{Neutrino and anti-neutrino fluxes at FASER$\nu$ predicted by the SM for an integrated luminosity of $\SI{150}{fb^{-1}}$. This corresponds to the luminosity expected in Run~3 of the LHC, at a collider center-of-mass energy of \SI{14}{TeV}. The kaon curves correspond to summed (anti-)neutrino fluxes from $K^\pm$, $K_L$ and $K_S$ decays.  Figure based on \cite{Kling:privcomm,Kling:2021gos}.}
  \label{fig:FASERflux}
\end{figure}

In this section we explain our procedure for estimating the sensitivity of FASER$\nu$ to new physics encapsulated in the Wilson coefficients $\epsilon_X$ of the WEFT Lagrangian \cref{eq:EFT_lweft}.
The master formula, \cref{eq:masterRate}, for calculating the FASER$\nu$ event rate requires as input the number of target nuclei ($N_T$), the SM neutrino fluxes at production ($\Phi_\alpha^{S,\rm SM}$), the SM neutrino detection cross section on the target nucleus ($\sigma_\beta^{\rm SM}$), and the modified oscillation probability ($\tilde P^S_{\alpha\beta}$).
The number of tungsten nuclei in FASER$\nu$ is $N_T = \num{3.14e27}$, which corresponds to a fiducial mass of \SI{0.96}{tons}. Fiducialization takes into account a geometrical acceptance of $80\%$ in order to suppress backgrounds~\cite{Abreu:2019yak}. 

We use neutrino fluxes that were kindly provided to us by the FASER$\nu$ collaboration \cite{Kling:privcomm} and are shown in \cref{fig:FASERflux}. They correspond to the fluxes from ref.~\cite{Kling:2021gos}, see also ref.~\cite{Bai:2020ukz}.
The event generator used was SIBYLL~2.3c~\cite{Ahn:2009wx,Riehn:2015oba,Riehn:2017mfm,Fedynitch:2018cbl}.
The detection cross section is given by \cref{eq:SMxsection-nu} for neutrinos and by \cref{eq:SMxsection-nubar} for anti-neutrinos.
We sum over the contributions from scattering on protons and on neutrons, taking into account the fact that tungsten has $n_p = 74$ protons and on average $n_n \simeq 110$ neutrons.  
In our formalism, all dependence on new physics is contained in the modified transition probability of \cref{eq:tildeP-L0}. 
Up to quadratic order in the Wilson coefficients it takes the form 
\begin{align}
 \tilde P^S_{\alpha\beta} (E_\nu)_{L=0}
    \simeq &   \ \delta_{\alpha\beta} 
       + 2 \sum_{X,j,k} p_{XL,\alpha}^{S,jk}  [\epsilon_X^{jk}]_{\alpha\alpha}
       \delta_{\alpha\beta}
        + 2 \sum_{j,k} d_{LL,\alpha}^{jk}  [\epsilon_L^{jk}]_{\beta \beta}
       \delta_{\alpha\beta}
       \nnl  + &
       \sum_{X,Y,j,k} \Big[ p_{XY,\alpha}^{S,jk} [\epsilon_X^{jk}]_{\alpha\beta} [\epsilon_Y^{jk}]_{\alpha\beta}
                           +  d_{XY,\beta}^{jk}    [\epsilon_X^{jk}]_{\beta\alpha}   [\epsilon_Y^{jk}]_{\beta\alpha} \Big].
    \label{eq:tildeP-simplified}
\end{align}
Here, we have taken into account that the detection coefficients $d_{LX}$ vanish for $X=R,S,P,T$, and that for all processes we consider here the production and detection coefficients are real.
By construction, the Wilson coefficients $[\epsilon_X^{jk}]_{\alpha\beta}$ that we want to constrain enter the experimental count rate only through \cref{eq:tildeP-simplified}. The production and detection coefficients have already been computed in~\cref{sec:Formalism}.  
The terms cubic and quartic in $\epsilon_X$, which are omitted from \cref{eq:tildeP-simplified} (but kept in our analysis), are relevant only when $\epsilon_X \gtrsim 1$.

In the SM limit, where all $\epsilon_X$ are zero, we recover $\tilde{P}_{\alpha\beta} = \delta_{\alpha\beta}$. 
At linear order in the new physics couplings, FASER$\nu$ is sensitive only to the lepton flavor-diagonal Wilson coefficients $[\epsilon_{X}^{jk}]_{\alpha\alpha}$. 
In fact, the linear terms are due to new physics modifying the partial decay widths of the source mesons into neutrinos and the detection cross-section. 
As these observables can be measured more precisely in dedicated (non-neutrino) precision experiments, we do not expect that the  FASER$\nu$ sensitivity to the linear terms can be competitive.
The situation changes at the quadratic level in  $\epsilon_X$, where the transition probability is no longer proportional to $\delta_{\alpha\beta}$. 
The advantage of FASER$\nu$ over non-neutrino precision experiments is here that it can identify the flavor of the emitted neutrino, and thus gain better access to the off-diagonal elements of the $[\epsilon_{X}^{jk}]_{\alpha\beta}$ matrices.
In particular, FASER$\nu$'s sensitivity to the $[\epsilon_{X}^{jk}]_{e\tau}$ and $[\epsilon_{X}^{jk}]_{\mu\tau}$ couplings is greatly enhanced by the large ratio of $\nu_{e,\mu}$ over $\nu_\tau$ fluxes. Thus, even a tiny fraction of $\nu_\tau$ from a production process that in the SM produces only $\nu_e$ and $\nu_\mu$ can be easily detected. Similarly, the sensitivity to even a small amount of anomalous $\tau$ lepton production in the detector induced by $\nu_e$ or $\nu_\mu$ is excellent, which leads to an enhanced sensitivity of FASER$\nu$ to $[\epsilon_{X}^{jk}]_{\tau e}$ and $[\epsilon_{X}^{jk}]_{\tau \mu}$.

As an example, consider the effects of the Wilson coefficients $\epsilon_R^{ud}$ on the number of tau events measured in FASER$\nu$. 
We find 
\begin{align}
    N_\tau \simeq  N_\tau^{\rm SM} \Big[ 1 + 0.25 \, [\epsilon^{ud}_R]_{e \tau}^2
                                           + \num{100} \, [\epsilon^{ud}_R]_{\mu \tau}^2
                                           + 40 \, [\epsilon^{ud}_R]_{\tau e}^2
                                           + \num{180} \, [\epsilon^{ud}_R]_{\tau \mu}^2 
                                           + \num{0.87} \, [\epsilon^{ud}_R]_{\tau \tau}^2 \Big] \,, 
\end{align}
where $N_\tau^{\rm SM} \simeq 17$ before applying acceptance and efficiency factors, and $N_\tau^{\rm SM} \simeq 10$ with these factors included. 
The sensitivity to $[\epsilon^{ud}_R]_{\tau e}$ and $[\epsilon^{ud}_R]_{\tau \mu}$ comes from the detection side, when the $\tau$ lepton is produced in the detector from incident $\nu_e$ or $\nu_\mu$, respectively. 
Even though the detection coefficients are only $\mathcal{O}(1)$ ($d_{RR}^{ud} \approx 0.5$ for neutrinos and  $\bar d_{RR}^{ud} \approx 1.6$ for anti-neutrinos), the large incident flux of $\nu_e$ and $\nu_\mu$ still leads to a sizeable event rate.
The effect is 5 times larger for $[\epsilon^{ud}_R]_{\tau \mu}$ than for $[\epsilon^{ud}_R]_{\tau e}$ simply because the incident flux of muon neutrinos is that much larger than the flux of electron neutrinos. 
The sensitivity to $[\epsilon^{ud}_R]_{e\tau}$ and $[\epsilon^{ud}_R]_{\mu\tau}$ comes from the production side from the decays $\pi \to e \nu_\tau$ or  $\pi \to \mu \nu_\tau$. 
In the presence of non-zero $[\epsilon^{ud}_R]_{\mu \tau}$, the large number of pion sources leads to sizeable anomalous production of $\nu_\tau$, albeit the effect is a bit smaller than for new physics on the detection side because $[\epsilon^{ud}_R]_{\mu \tau}$ does not affect neutrinos from kaon decays.
By a similar argument, we also expect that the sensitivity to $[\epsilon^{ud}_R]_{e \tau}$ will be very poor. While the flux of electron neutrinos in the SM is sizeable, that flux is entirely dominated by kaon, charm, and hyperon decays (see \cref{fig:FASERflux}), which are unaffected by $[\epsilon^{ud}_R]_{e \tau}$. Only $\pi \to e \nu$ decays are enhanced, but their strong chiral suppression precludes any sizeable increase in the overall count rate at FASER$\nu$.
This simple example gives a qualitative understanding of the sensitivity to various Wilson coefficients, which in the following is estimated by a more elaborate analysis.    

We fold the event rate with a Gaussian energy smearing function with a width of $0.3 E_\nu$~\cite{Abreu:2019yak}, and we then apply the vertex reconstruction efficiency taken from fig.~9 of ref.~\cite{Abreu:2019yak} as well as the charged lepton identification efficiency, which is close to $\varepsilon_e = 100\%$ for electrons, $\varepsilon_\mu = 86\%$ for muons, and $\varepsilon_\tau = 75\%$ for taus, see sec.~VI.C of ref.~\cite{Abreu:2019yak}. Moreover, note that the differences in acceptance between 2-body and 3-body kaon decays are already accounted for by the factors $\beta^S_i(E_S)$ in the production coefficients in \cref{eq:pxy}. The main sources of background at FASER$\nu$ are muons produced at the ATLAS interaction point or further downstream, as well as secondary particles that could mimic the neutrino signals. However, these backgrounds can be suppressed to a negligible level by the fiducialization cut, by only considering reconstructed neutrino energies $E_\nu > \SI{100}{GeV}$, and by additional kinematic cuts whose effect on the signal is encoded in the above-mentioned efficiency factors. This last set of cuts, discussed in detail in sec.~V.C of ref.~\cite{Abreu:2019yak}, includes for instance a cut on the total momentum fraction carried by the highest momentum particle, as well as a cut on its angle with respect to the other particles in the event.

\begin{figure}
  \centering
  \includegraphics[width=1\textwidth]{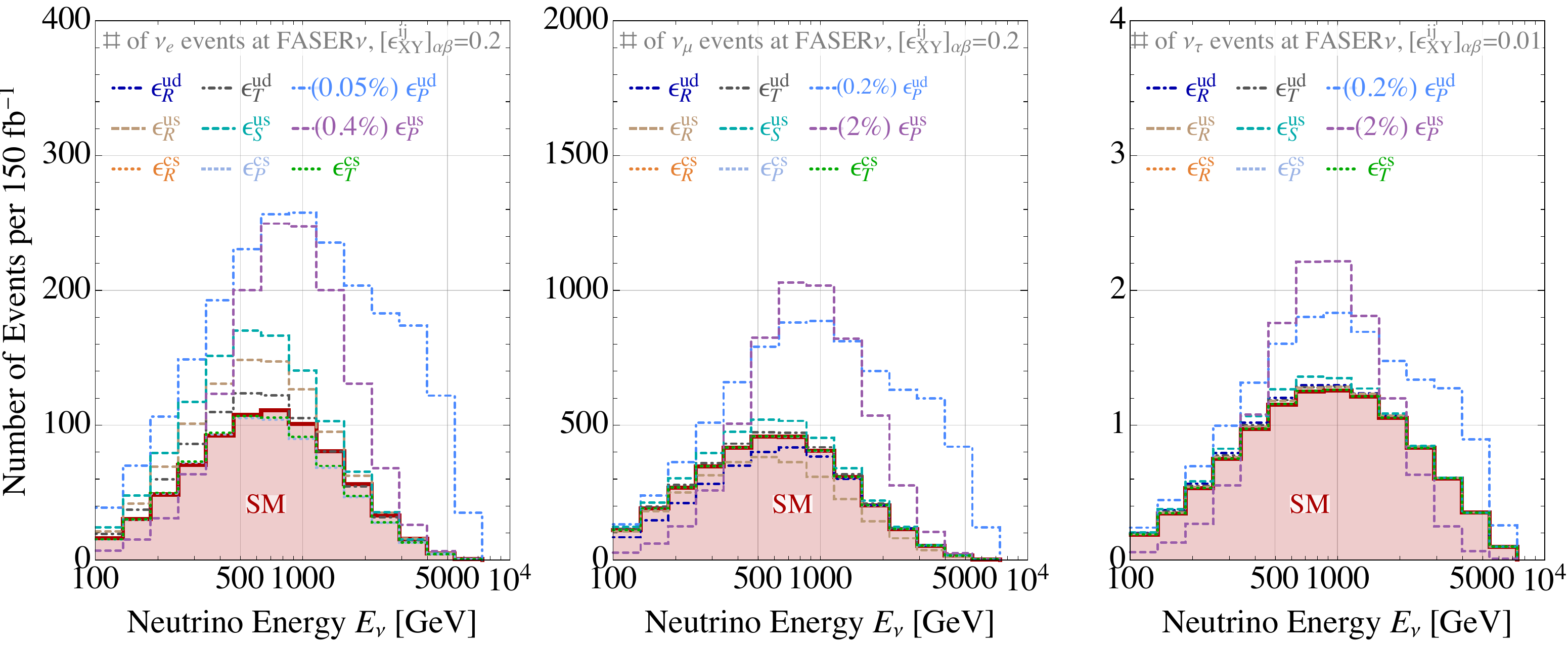}
  \caption{Predicted event spectra at FASER$\nu$. All the $\epsilon$'s for $\nu_e$ and $\nu_\mu$ are set to $0.2$, while they are $0.01$ for the $\nu_\tau$ events. For pseudoscalar couplings we rescale the number of events by the indicated factor.}
  \label{fig:spectrum}
\end{figure}

In our analysis, we consider neutrino energies $\SI{e2}{GeV} \leq E_\nu \leq \SI{e4}{GeV}$, sorted into 15 log-spaced bins for $\nu_e$ and $\nu_\mu$, and combined into a single bin for $\nu_\tau$.
With the efficiencies taken into account, we predict that, without new physics, FASER$\nu$ will detect about $n_{e} + n_{\bar{e}} = \num{908}$ electron neutrinos and anti-neutrinos, $n_{\mu} + n_{\bar{\mu}} = 4979$ muon neutrinos and anti-neutrinos, and $n_{\tau} + n_{\bar{\tau}} = 17$ tau neutrinos and anti-neutrinos. These numbers differ from the ones shown in Table~II of ref.~\cite{Abreu:2019yak} because we are using updated neutrino fluxes~\cite{Kling:privcomm,Kling:2021gos} compared to that reference. We have checked that, using instead the neutrino fluxes from ref.~\cite{Abreu:2019yak}, we do reproduce the event numbers from that paper.
We plot the expected event spectra for the different neutrino flavors in \cref{fig:spectrum}.

To investigate the sensitivity of FASER$\nu$ to new physics, we define the Gaussian log-likelihood function 
\begin{align}
  \chi^2(\epsilon_X)
    &= \sum_{\nu,\bar\nu} \sum_{\beta=e,\mu,\tau} \sum_i
       \frac{\big[ N_\beta^i(\vec{a}, \epsilon_X) - N_\beta^{\text{SM},i} \big]^2}
            {N_\beta^i(\vec{a}, \epsilon_X)}
     + \sum_{\alpha=e,\mu,\tau} \frac{a_\alpha^2}{\sigma_\alpha^2} \,,
  \label{eq:chisq}
\end{align}
where the number of $\nu_\beta$-like events in the $i$-th energy bin is given by $N_\beta^{\text{SM},i}$ for the SM, and by $N_\beta^i(\vec{a}, \epsilon_X)$ in the presence of new physics. The latter quantity depends not only on the Wilson coefficients $\epsilon_X$, but also on a set of nuisance parameters $a_\alpha$ ($\alpha = e, \mu, \tau$) which parameterize a systematic normalization bias in the primary meson and SM neutrino fluxes. More precisely, based on \cref{eq:masterRate}, $N_\beta^i(\vec{a}, \epsilon_X)$ is given by
\begin{align}
    N_\beta^i(\vec{a}, \epsilon_X)
      = N_T \int_{E^i_\text{min}}^{E^i_\text{max}} \! dE_\nu \,
        \sigma^{\text{SM}}_\beta(E_\nu) \,
        \sum_{\alpha, S} (1 + a_\alpha) \Phi^{S,\text{SM}}_\alpha(E_\nu) \, 
        \tilde{P}^S_{\alpha\beta}(E_\nu,L) \,,
\end{align}
where $E^i_\text{min}$ and $E^i_\text{max}$ denote the boundaries of the $i$-th energy bin. For the uncertainties of the nuisance parameters, we will use two sets of values, namely a more optimistic one with $\sigma_e=5\%$, $\sigma_\mu=10\%$, and $\sigma_\tau=15\%$ for electron, muon, and tau (anti-)neutrinos, respectively (based on Table~II of ref.~\cite{Abreu:2019yak}), and a more conservative one with $\sigma_e=30\%$, $\sigma_\mu=40\%$, and $\sigma_\tau=50\%$.  Finally we sum over neutrinos and anti-neutrinos.

In computing the projected limits on the Wilson coefficients, we assume that FASER$\nu$ will observe exactly the number of events predicted by the SM in each flavor. We allow only one of the $[\epsilon_X^{jk}]_{\alpha\beta}$ coefficients to be non-zero at a time.

%=======================================================================
\section{Results}
\label{sec:results}
%=======================================================================

\begin{figure}
  \centering
  \includegraphics[width=0.95\textwidth]{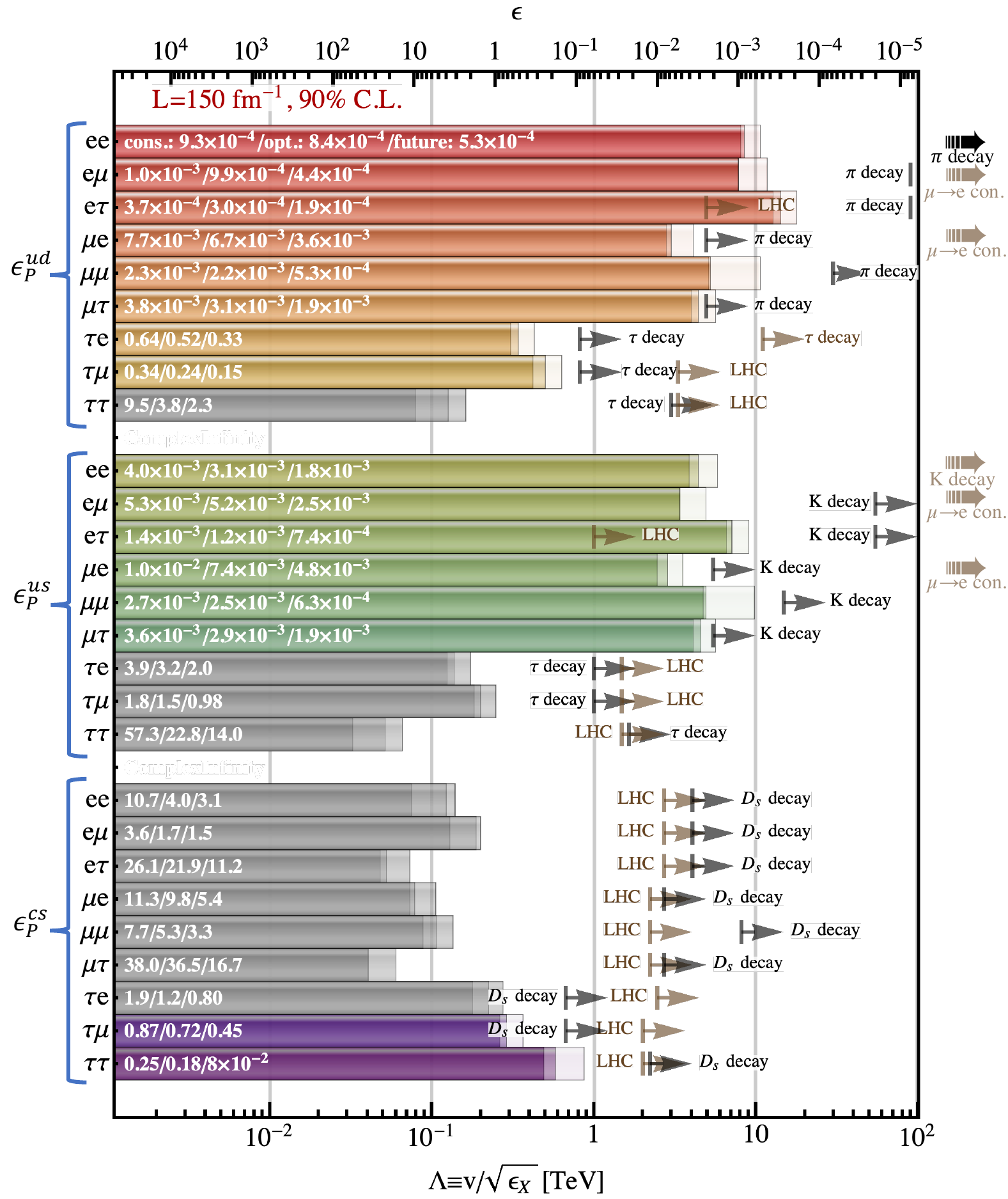}
  \caption{The projected FASER$\nu$ constraints on the Wilson coefficients of new pseudoscalar interactions in the Weak Effective Field Theory~(WEFT) framework. Projected limits are reported both in terms of the dimensionless couplings $[\epsilon_X^{jk}]_{\alpha\beta}$ (top axis) and in terms of the effective new physics scale $[\Lambda_X^{jk}]_{\alpha\beta} \equiv v / \sqrt{[\epsilon_X^{jk}]_{\alpha\beta}}$ (bottom axis). Each colored bar indicates the limit on one particular interaction, with the shorter, darker piece corresponding to conservative assumptions on the systematic uncertainties, the lighter colored piece corresponding to more optimistic assumptions, and the unshaded extension on the right indicating the sensitivity of FASER$\nu$ during the high-luminosity LHC phase, with a factor of 20 more statistics.
  }
  \label{fig:constraints-summary-P}
\end{figure}

\begin{figure}
  \centering
  \includegraphics[width=.95\textwidth]{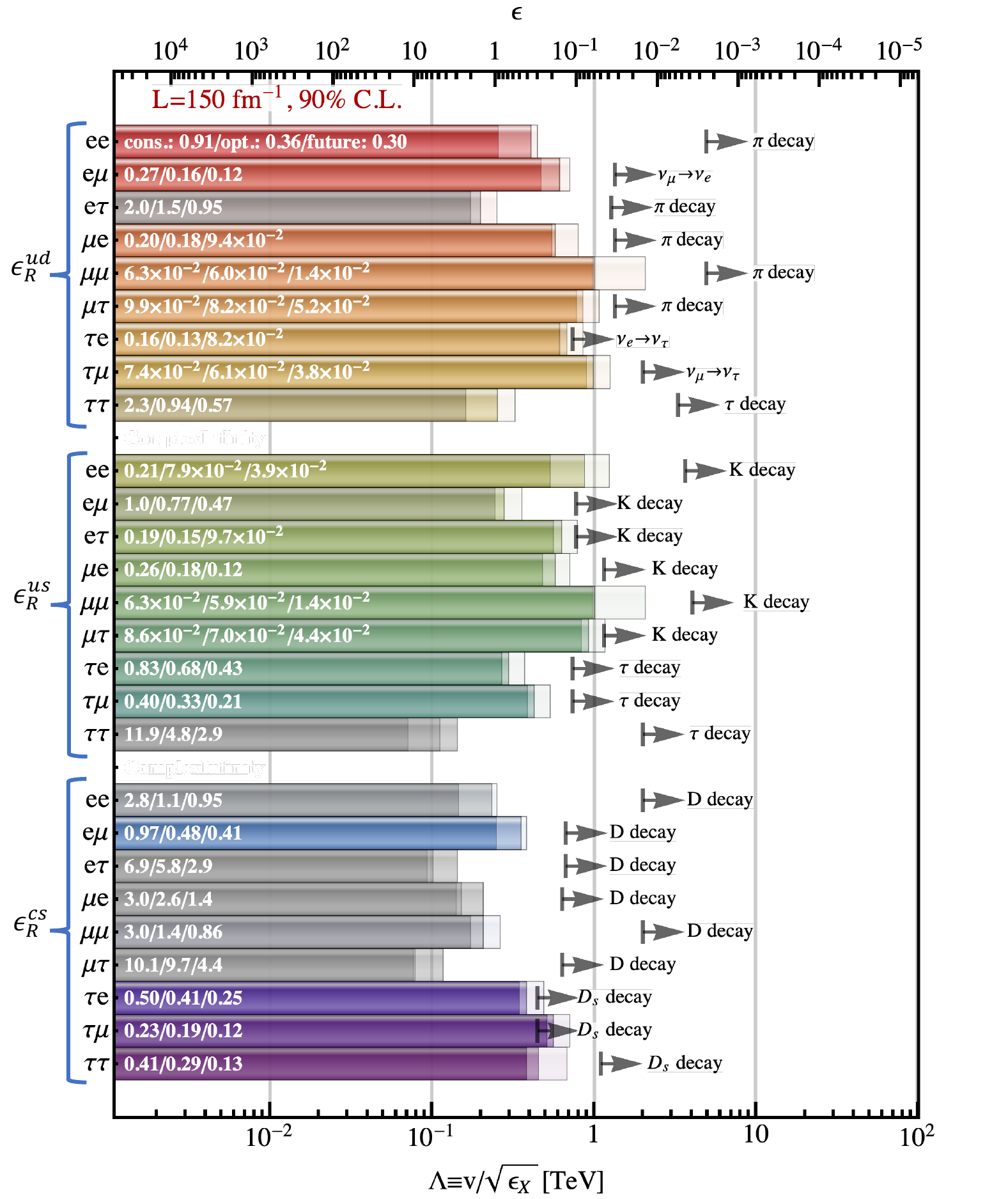}
  \caption{Similar to \cref{fig:constraints-summary-P}, but showing constraints on right-handed couplings.}
  \label{fig:constraints-summary-R}
\end{figure}

\begin{figure}
  \centering
  \includegraphics[width=.95\textwidth]{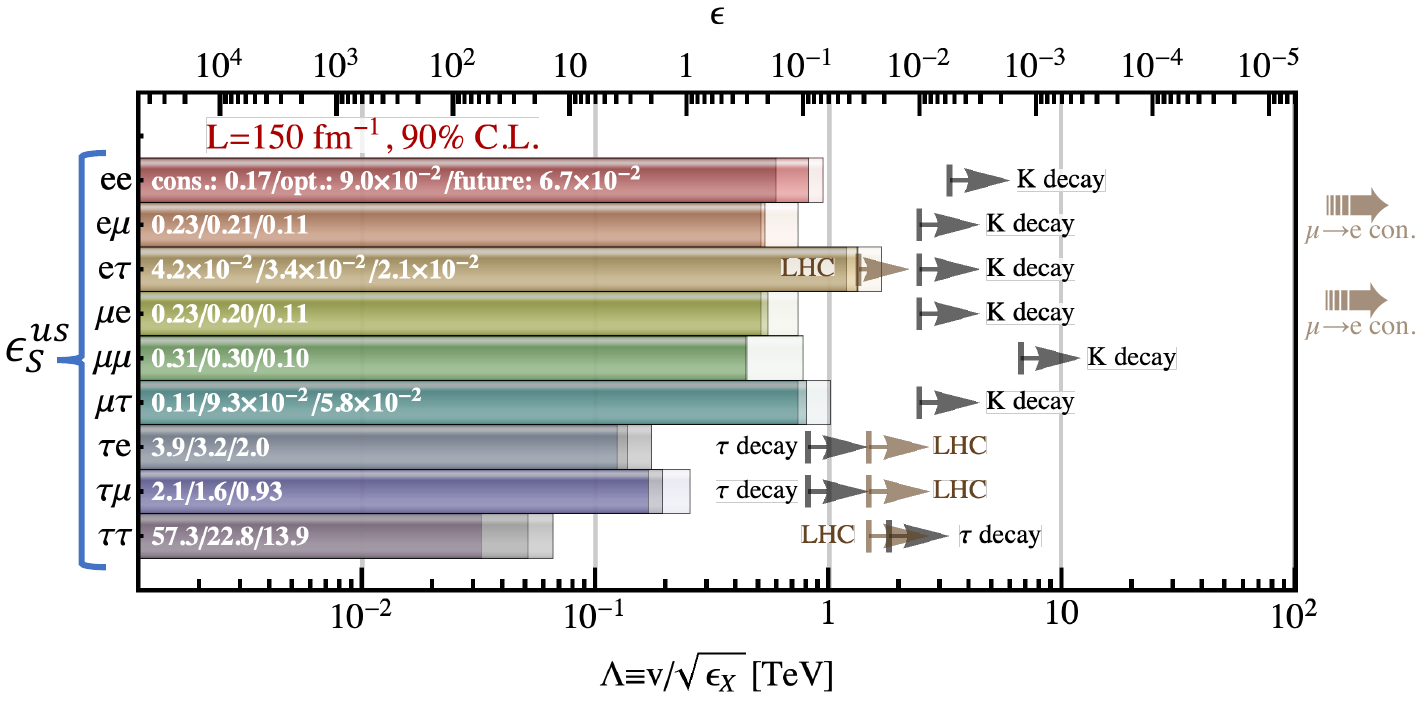}
  \caption{Similar to \cref{fig:constraints-summary-P}, but showing constraints on scalar couplings.}
  \label{fig:constraints-summary-S}
\end{figure}

\begin{figure}
  \centering
  \includegraphics[width=.95\textwidth]{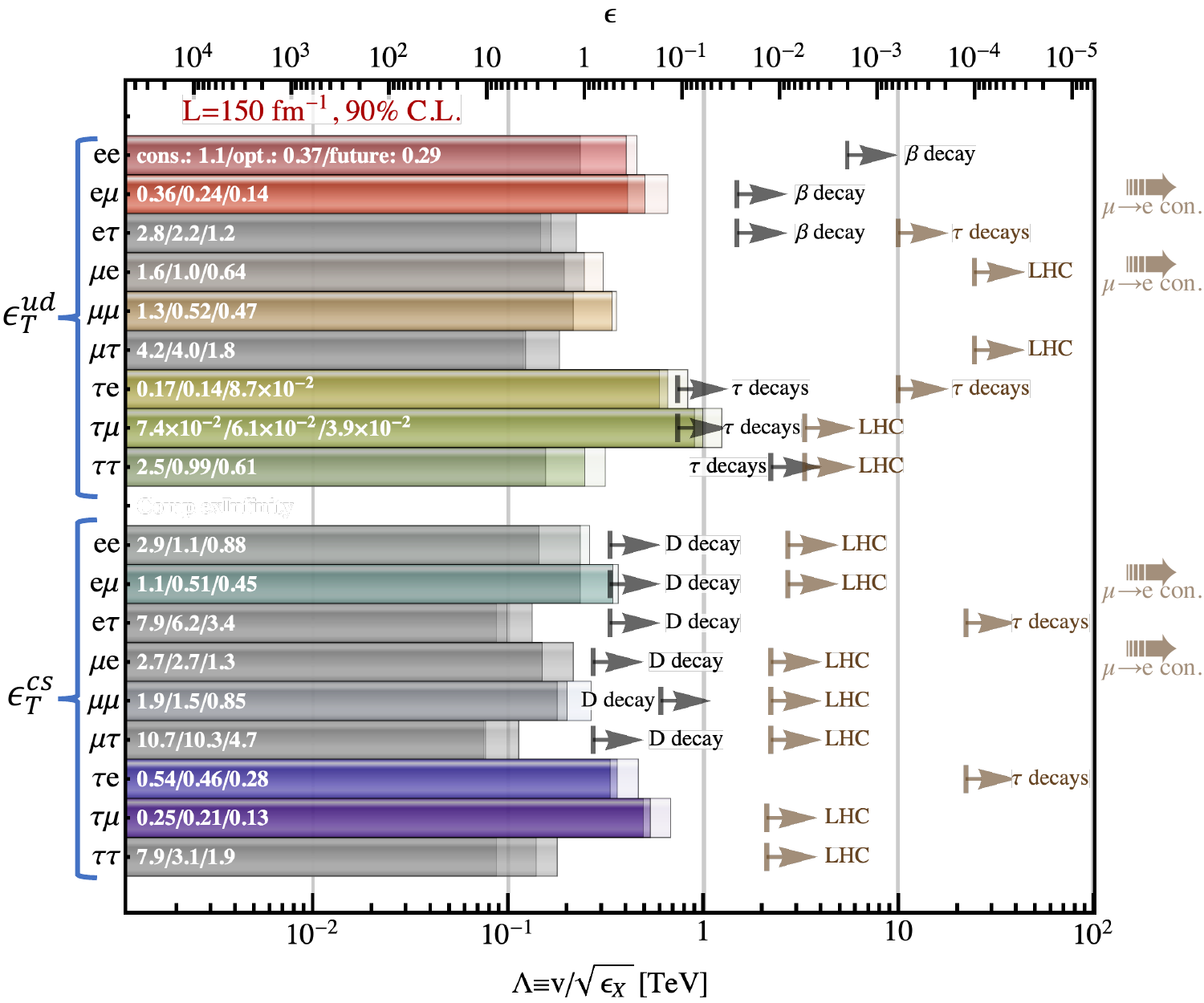}
  \caption{Similar to \cref{fig:constraints-summary-P}, but showing constraints on tensor couplings.}
  \label{fig:constraints-summary-T}
\end{figure}

We now proceed to the discussion of our main results, namely the projected constraints on the Wilson coefficients appearing in \cref{eq:EFT_lweft}. 
We summarize these constraints in \cref{fig:constraints-summary-R,fig:constraints-summary-S,fig:constraints-summary-P,fig:constraints-summary-T} for right-handed, scalar, pseudoscalar, and tensor couplings, respectively. 
In addition to quoting limits in terms of the dimensionless $[\epsilon_X^{jk}]_{\alpha\beta}$ parameters (top axis and labels inside each bar), we also express them in terms of the effective new physics scale $[\Lambda_X^{jk}]_{\alpha\beta} \equiv v / ([\epsilon_X^{jk}]_{\alpha\beta})^{1/2}$.
Operators for which $\epsilon > 1$ is required to saturate the limit are shown in gray to emphasize that, for such large couplings, our formalism is expected to become less accurate due to threshold effects as the cutoff scale of the theory approaches the center-of-mass energy of neutrino scattering in FASER$\nu$. 
We do not consider WEFT corrections to left-handed interactions in our analysis because the intricate interplay of these interactions with SM processes implies that deriving a reliable constraint would require re-extracting the CKM elements from the data, taking into account possible contamination by new physics. As mentioned already in \cref{sec:FASER}, we allow only one of the $[\epsilon_X^{jk}]_{\alpha\beta}$ coefficients to be non-zero entries at a time. Each row of \cref{fig:constraints-summary-R,fig:constraints-summary-S,fig:constraints-summary-P,fig:constraints-summary-T} shows three overlapping bars, the leftmost one corresponding to very conservative systematic uncertainties, the middle one corresponding to more optimistic systematic uncertainties (see \cref{sec:FASER}), and the rightmost one indicating the sensitivity of FASER$\nu$ during the high-luminosity LHC phase, with \SI{3}{ab^{-1}} of integrated luminosity. Further improvements could of course be expected if also the detector were to be enlarged.

Starting with constraints on pseudoscalar couplings, we see from \cref{fig:constraints-summary-P} that FASER$\nu$ will be able to constrain many of the entries of the $\epsilon_P^{ud}$ and $\epsilon_P^{us}$ matrices at the per mille level (see top and middle parts of \cref{fig:constraints-summary-P}).
The reason for FASER$\nu$'s excellent sensitivity to these couplings is the strong chiral enhancement of the production coefficients for fully leptonic meson decays, see e.g.\ \cref{eq:pioncoefficients} for the case of $p^{\pi,ud}_{PL,\mu}$ and $p^{\pi,ud}_{PP,\mu}$.
The impact of this enhancement may not be immediately obvious for the decays $\pi^+ \to \mu^+ \nu_\mu$ and $K^+ \to \mu^+ \nu_\mu$, which are already dominant even in the SM.
As an already large branching ratio cannot be enhanced much further by new physics, one might expect that lifting chiral suppression would not change the $\nu_\mu$ flux by a lot.
However, the presence of new physics changes also the total meson decay rate, and thus the fraction of mesons that can decay before being stopped in the matter they encounter downstream of the ATLAS interaction point (beam pipe, tunnel walls, etc.).
Effectively, by detecting neutrinos from dominant decays like $\pi^+ \to \mu^+ \nu_\mu$ and $K^+ \to \mu^+ \nu_\mu$, FASER$\nu$ is carrying out a precision measurement of the pion and kaon partial decay widths into a specific neutrino flavor.
Comparing this measurement to a first-principles prediction (based on lattice QCD results for the decay constant) is what allows the experiment to set limits on new physics even when the new physics affects the leading meson decay modes.
For new physics operators that affect only sub-leading decays, one can typically achieve smaller theory errors because no first-principles prediction of a total decay rate is necessary -- the latter is fixed in-situ by measuring the dominant decay mode.
All that is needed in this case is a prediction for the branching ratio of the sub-dominant decay channel.
Chiral enhancement is even stronger for charged meson decays to electrons, such that the decay $\pi^+ \to e^+ \nu_\alpha$ is enhanced to the extent that the constraints on $[\epsilon_P^{ud}]_{e\alpha}$ are as strong as the ones on $[\epsilon_P^{ud}]_{\mu\alpha}$.

Constraints on couplings to light flavor quarks involving $\tau$ leptons ($[\epsilon_P^{ud}]_{\tau\alpha}$ and $[\epsilon_P^{us}]_{\tau\alpha}$) depend entirely on the detection process because pions and kaons cannot decay to $\tau$'s. 
These constraints are therefore generally weaker than those on couplings to electrons and muons.
They are weaker for $[\epsilon_P^{us}]_{\tau\alpha}$ than for $[\epsilon_P^{ud}]_{\tau\alpha}$ because neutrino interactions through the former are in addition suppressed by the CKM element $V_{us}$ (see our definition of $[\epsilon_X^{us}]_{\tau\beta}$ in \cref{eq:EFT_lweft}) and by the strange quark PDF. (Anti-neutrinos can still interact on valence up quarks, but their flux is slightly lower.)

Chiral enhancement is weak in the case of fully leptonic charm decays because of the appearance of $m_c$ in the denominators of \cref{eq:Dscoefficients}.
As the fully leptonic branching ratio of charm mesons is very small in the SM, the enhancement is not sufficient to significantly increase the associated neutrino flux, therefore constraints on $\epsilon_P^{cs}$ (bottom part of \cref{fig:constraints-summary-P}) are generally weaker than those on $\epsilon_P^{ud}$ and $\epsilon_P^{us}$. The exception are couplings involving $\tau$ leptons because, unlike pions and kaons, charm mesons can decay to $\tau$s.

Comparing FASER$\nu$'s sensitivity to pseudoscalar new physics to the existing constraints that will be discussed in more detail in \cref{sec:comparison}, we see that for some couplings -- especially those benefiting from chiral enhancement -- FASER$\nu$ will be more sensitive than other LHC searches. On the other hand, precision measurements of meson decays in low-energy experiments will typically still have an edge over FASER$\nu$, even though for some operators future measurements with LHC neutrinos will be quite competitive.  Should a deviation from the SM prediction be found, an important aspect of FASER$\nu$ and other, similar, experiments will be their sensitivity to the neutrino flavor, an observable that meson decay measurements are insensitive to. This highlights the unique potential of LHC neutrinos in hunting for new physics and is one of the main conclusions of this paper.

\vspace{1em}
We now turn our attention to right-handed couplings involving first-generation quarks, $\epsilon_R^{ud}$.
We see from the top part of \cref{fig:constraints-summary-R} that FASER$\nu$'s sensitivity to new physics in this sector will be at the 10\% level. This is worse than the sensitivity to pseudoscalar interactions due to the lack of chiral enhancement. Nevertheless, some of the $\epsilon_R^{ud}$ benefit from other types of enhancement: for the diagonal couplings $[\epsilon_R^{ud}]_{ee}$ and $[\epsilon_R^{ud}]_{\mu\mu}$, interference with the SM amplitude implies sensitivity at the linear order. The off-diagonal coupling $[\epsilon_R^{ud}]_{\mu\tau}$ converts part of the very large muon (anti-)neutrino flux into $\nu_\tau$, for which the SM background flux is low.
Constraints on $[\epsilon_R^{ud}]_{\tau e}$ and $[\epsilon_R^{ud}]_{\tau\mu}$ are entirely based on detection processes in which a $\nu_e$ or $\nu_\mu$ creates a $\tau$ lepton. Once again, these processes benefit from the low SM rate in the $\tau$ channel. There is no contribution from the production side because pions (the only unflavored mesons we consider) cannot decay to $\tau$ leptons.

Considering next right-handed couplings to up and strange quarks (middle part of \cref{fig:constraints-summary-R}), we find fairly strong constraints for all lepton flavor structures thanks to the fact that the flux of forward kaons at the LHC is large, and that kaons have sizeable branching ratios into both muons and electrons. 
Only the limit on $[\epsilon_X^{us}]_{\tau\tau}$ is weaker than the others because it comes purely from modified detection processes, given that kaons cannot decay to $\tau$ leptons for kinematic reasons.

Constraints on right-handed couplings to charm and strange quarks (bottom part of \cref{fig:constraints-summary-R}), are in general weaker because the flux of charm mesons is lower than the one of pions and kaons, and because detection reactions sensitive to $\epsilon_X^{cs}$ are suppressed by sea quark PDFs.
Among the $[\epsilon_R^{cs}]_{\alpha\beta}$ coefficients, the most strongly constrained ones are $[\epsilon_R^{cs}]_{\tau e}$, $[\epsilon_R^{cs}]_{\tau\mu}$, and $[\epsilon_R^{cs}]_{\tau\tau}$. The first two of these lead to the production of $\tau$ leptons in the detector off the large $\nu_e$ and $\nu_\mu$ fluxes, while the latter one profits from sensitivity at the linear order.
Similarly, $[\epsilon_R^{cs}]_{e\mu}$ allows $\nu_\mu$ to produce electron-like signatures, and since the background of electron-like events is lower than the one for muon-like events, this leads to enhanced sensitivity.

In comparison to existing limits, we see that FASER$\nu$ constraints on $\epsilon_R^{jk}$ will not be quite competitive yet.
However, the difference in sensitivity is small in some cases and may be overcome by future upgrades to FASER with a larger detector (and hence larger acceptance) \cite{felix_kling_2020_4009641} and with an increased neutrino flux from the high-luminosity LHC.

\vspace{1em}
Looking next at constraints on scalar couplings (\cref{fig:constraints-summary-S}), we remark that we do not show expected limits on $\epsilon_S^{ud}$ and $\epsilon_S^{cs}$ here.
The reason is that the production coefficients for these couplings vanish in the case of pion and charm decays (see \cref{sec:production}), and the detection coefficients are small (see \cref{sec:detection}).
The sensitivity to $\epsilon_S^{ud}$ and $\epsilon_S^{cs}$ is therefore extremely poor.
For the case of $\epsilon_S^{us}$, on the other hand, we obtain decent limits thanks to the sizeable production coefficients in 3-body kaon decays. Nevertheless, these limits are not quite competitive with existing constraints from collider studies and precision measurements of kaon decays.

\vspace{1em}
Last but not least we have investigated FASER$\nu$'s potential to constrain tensor interactions parameterized by $\epsilon_T^{jk}$, see \cref{fig:constraints-summary-T}.
We first note that the production coefficients $p_{TY}^{ud}$, $p_{XT}^{ud}$, $p_{TY}^{cs}$, and $p_{XT}^{cs}$ vanish (see discussion in \cref{sec:production}), while the detection coefficients are sizeable (see \cref{tab:dxy}).
This leads to limits that are decent, but not competitive.
We see that here, the $e\mu$, $\tau e$, and $\tau\mu$ elements of $\epsilon_T^{ud}$ and $\epsilon_T^{cs}$ are the most constrained ones because they correspond to interactions in which a neutrino flavor with a sizeable flux creates charged leptons that in the SM can only be produced by a less abundant neutrino species.
This leads to an excellent signal-to-background ratio for these channels.
It is noteworthy in particular that the FASER$\nu$ limit on $[\epsilon_T^{ud}]_{\tau\mu}$ can potentially beat the one from $\tau$ decays (though not the one from top decays, see \cref{sec:comparison}). 
For couplings to up and strange quarks, the production coefficients do not vanish, but are still very small (see \cref{fig:pXY}), and the detection coefficients are CKM-suppressed.
Therefore, limits on $\epsilon_T^{us}$ are very weak and are not shown here.

%=======================================================================
\section{Comparison with other experiments}
\label{sec:comparison}
%=======================================================================

In \cref{fig:constraints-summary-P,fig:constraints-summary-R,fig:constraints-summary-S,fig:constraints-summary-T}, we have already compared FASER$\nu$'s sensitivity to new neutrino interactions with existing constraints from other experimental probes.
In the following, we explain how these external limits are obtained.
We will focus on bounds obtained in the framework of WEFT, that is bounds from low-energy experiments sensitive to charged-current interactions. For many couplings, we will also show constraints that are only valid if the UV-completion of WEFT is SMEFT. We will do so in particular when the bounds obtained in SMEFT are superior to the WEFT-only constraints.

The bounds from neutrino experiments, meson (semi-)leptonic decay and $\beta$-decays can be directly compared to the FASER$\nu$ projections as they are evaluated at an energy scale that is well captured by the WEFT. 
On the other hand, bounds from colliders and charged-lepton flavor violation are valid only under the assumption that WEFT is UV completed by SMEFT. 
Bounds are given at 90\% CL (unless otherwise stated), assuming only one operator is nonzero, and using a renormalization scale $\mu=\SI{2}{GeV}$ in the $\overline{\rm MS}$ scheme.
We assume all Wilson coefficients are real, and the bounds given in the following refer to their absolute value, $|[\epsilon^{jk}_X]_{\alpha\beta}|$.
We collect the strongest bound, to our knowledge, for each Wilson coefficient in \cref{tab:external-constraints-P},~\ref{tab:external-constraints-R} and~\ref{tab:external-constraints-ST}. Entries printed in bold face in these tables have been derived in this work, while those printed with a normal font weight are taken from the literature.

%=======================================================================
\sisetup{detect-weight=true,detect-inline-weight=math}
\begin{table}[t]
    \centering
    \scalebox{0.8}{
    \begin{tabular}{l|cc|cc}
        \toprule
        Coupling 
        &  \multicolumn{2}{c|}{Low energy (WEFT)} 
        & \multicolumn{2}{c}{High energy / CLFV (SMEFT)} \\
        & 90\,\% CL bound & process & 90\,\% CL bound & process\\
        \midrule
        $[\epsilon_P^{ud}]_{ee}$       
        &\bfseries\num{4.6e-7}  % MGA and JK
        & $\bf{\Gamma_{\pi\to e\nu} / \Gamma_{\pi\to\mu\nu}}$
        &
        &
        \\
        $[\epsilon_P^{ud}]_{e\mu}$     
        & \num{7.3e-6}  % MGA and JK
        & $\Gamma_{\pi\to e\nu} / \Gamma_{\pi\to\mu\nu}$~\cite{Falkowski:2019xoe} 
        &  \bf{\num{2.0e-8}} 
        &   \bf{$\mu \to e$ conversion}
        \\
        $[\epsilon_P^{ud}]_{e\tau}$     
        & \num{7.3e-6}  % MGA and JK
        & $\Gamma_{\pi\to e\nu} / \Gamma_{\pi\to\mu\nu}$~\cite{Falkowski:2019xoe}
        & {\num{2.5e-3}}
        & LHC~\cite{Cirigliano:2021img}
        \\
        $[\epsilon_P^{ud}]_{\mu e}$    
        &  \bfseries\num{2.6e-3} % MGA and JK
        & $\bf{\Gamma_{\pi\to e\nu} / \Gamma_{\pi\to\mu\nu}}$
        & \bf{\num{2.0e-8}} 
        &  \bf{$\mu \to e$ conversion}
        \\%~\cite{XYZ} \\
        $[\epsilon_P^{ud}]_{\mu\mu}$   
        & \bfseries\num{9.4e-5} % MGA and JK
        & $\bf{\Gamma_{\pi\to e\nu} / \Gamma_{\pi\to\mu\nu}}$\\
        $[\epsilon_P^{ud}]_{ \mu\tau}$    
        &  \bfseries\num{2.6e-3} % MGA and JK
        & $\bf{\Gamma_{\pi\to e\nu} / \Gamma_{\pi\to\mu\nu}}$ \\
        $[\epsilon_P^{ud}]_{\tau e}$   
        & \bfseries{\num{9.0e-2}}  % AF and JK
        & $\bf{\Gamma_{\tau \to \pi \nu}}$   % $\tau$-decay 
        & \num{5.8e-3}${}^{(*)}$/\num{4.4e-4} % AF and YS
        & LHC~\cite{Cirigliano:2018dyk} / $\tau$ decay~\cite{Cirigliano:2021img} 
        \\
        $[\epsilon_P^{ud}]_{\tau\mu}$   
        & \bfseries{\num{9.0e-2}}  % AF and JK
        & $\bf{\Gamma_{\tau \to \pi \nu}}$  
        & \num{5.8e-3}${}^{(*)}$ % AF and YS
        & LHC~\cite{Cirigliano:2018dyk}\\
        $[\epsilon_P^{ud}]_{\tau\tau}$ 
        & \num{8.4e-3} 
        & $\tau$-decay~\cite{Cirigliano:2018dyk}
        & \num{5.8e-3}${}^{(*)}$ % AF and YS
        & LHC~\cite{Cirigliano:2018dyk} \\
        \midrule
        %------------------------------------------------------------
        $[\epsilon_P^{us}]_{ee}$       
        & \bfseries{\num{1.1e-6}}   % MGA and JK
        & $\bf{\Gamma_{K\to e\nu}/\Gamma_{K\to\mu\nu}}$ 
        \\
        $[\epsilon_P^{us}]_{e\mu}$     
        & \bfseries{\num{2.1e-5}}   % MGA and JK
        & $\bf{\Gamma_{K\to e\nu}/\Gamma_{K\to\mu\nu}}$
        & \bf{\num{6.2e-7}} 
        &  \bf{$\mu \to e$ conversion}
        \\
        $[\epsilon_P^{us}]_{e\tau}$     
        & \bfseries{\num{2.1e-5}}   % MGA and JK
        & $\bf{\Gamma_{K\to e\nu}/\Gamma_{K\to\mu\nu}}$ 
        & \num{7.1e-2}
        & LHC~\cite{Cirigliano:2021img}\\
        $[\epsilon_P^{us}]_{\mu e}$  
        & \bfseries{\num{2.3e-3}}  % MGA and JK
        & $\bf{\Gamma_{K\to e\nu}/\Gamma_{K\to\mu\nu}}$
        & \bf{\num{6.2e-7}}  
        &  \bf{$\mu \to e$ conversion}
        \\%~\cite{XYZ}  \\
        $[\epsilon_P^{us}]_{\mu\mu}$   
        & \bfseries{\num{2.2e-4}}   % MGA and JK
        & $\bf{\Gamma_{K\to e\nu}/\Gamma_{K\to\mu\nu}}$
        \\%~\cite{XYZ} \\
        $[\epsilon_P^{us}]_{\mu\tau}$  
        & \bfseries{\num{2.3e-3}}  % MGA and JK
        & $\bf{\Gamma_{K\to e\nu}/\Gamma_{K\to\mu\nu}}$
        \\%~\cite{XYZ}  \\
        $[\epsilon_P^{us}]_{\tau e}$  
        &  \bfseries{\num{6.4e-2}}  % AF and JK
        & $\bf{\Gamma_{\tau \to K \nu}/\Gamma_{K \to \mu \nu}}$ 
        & {\bfseries{\num{3.1e-2}${}^{(*)}$}}/\num{8.1e-2}  % JK and AF for LHC
        &{\bf{ LHC (data~\cite{Aaboud:2018vgh})}}/$\tau$-decay~\cite{Cirigliano:2021img}\\
        $[\epsilon_P^{us}]_{\tau\mu}$  
        & \bfseries{\num{6.4e-2}}  % AF and JK
        & {$\bf{\Gamma_{\tau \to K \nu}/\Gamma_{K \to \mu \nu}}$} 
        &{\bfseries{\num{3.1e-2}${}^{(*)}$}}  % JK and AF
        &{\bf{ LHC (data~\cite{Aaboud:2018vgh})}}\\
        $[\epsilon_P^{us}]_{\tau\tau}$ 
        & \num{1.3e-2} % MGA, AF and JK
        & $\tau$-decay~\cite{Gonzalez-Solis:2019owk} % eq. 13
        & {\bfseries{\num{3.1e-2}${}^{(*)}$}} % JK and AF
        & {\bf{ LHC (data~\cite{Aaboud:2018vgh})}}\\
        \midrule
        %------------------------------------------------------------
        $[\epsilon_P^{cs}]_{ee}$      
        & {$\bf4.8 \times 10^{-3}$}  % AF and JK
        & {$\bf{\Gamma_{D_s \to e \nu}}$}
        & \num{1.3e-2} 
        & LHC~\cite{Fuentes-Martin:2020lea}\\
        $[\epsilon_P^{cs}]_{e\mu}$     
        & {$\bf4.6 \times 10^{-3}$} % AF and JK
        & {$\bf{\Gamma_{D_s \to e \nu}}$ }
        & \num{1.3e-2} / {\bf\num{2.7e-6}} 
        & LHC~\cite{Fuentes-Martin:2020lea} / {\bf $\mu \to e$ conversion} \\
        $[\epsilon_P^{cs}]_{e\tau}$     
        & {$\bf4.6 \times 10^{-3}$} % AF and JK
        & $\bf{\Gamma_{D_s \to e \nu}}$
        & \num{1.3e-2} 
        / \num{1.9e-2} 
        & LHC / $\tau$-decays~\cite{Fuentes-Martin:2020lea,Cirigliano:2021img}\\
        $[\epsilon_P^{cs}]_{\mu e}$  
        & {$\bf8.9 \times 10^{-3}$ } 
        & $\bf{\Gamma_{D_s \to \mu \nu}}$ 
        & \num{2.0e-2} / {\bf\num{2.7e-6}}
        & LHC~\cite{Fuentes-Martin:2020lea} / {\bf $\mu \to e$ conversion} \\
        $[\epsilon_P^{cs}]_{\mu\mu}$   
        & {$\bf1.0 \times 10^{-3}$}  % AF and JK
        & $\bf{\Gamma_{D_s \to \mu \nu}}$ 
        & \num{2.0e-2} 
        & LHC~\cite{Fuentes-Martin:2020lea}\\
        $[\epsilon_P^{cs}]_{\mu\tau}$  
        & {$\bf8.9 \times 10^{-3}$ } 
        & $\bf{\Gamma_{D_s \to \mu \nu}}$ 
        & \num{2.0e-2} 
        & LHC~\cite{Fuentes-Martin:2020lea}\\
        $[\epsilon_P^{cs}]_{\tau e}$  
        & {$\bf2.0 \times 10^{-1}$} % AF and JK
        &$\bf{\Gamma_{D_s \to \tau \nu}}$
        & \num{1.6e-2} / \num{1.9e-2}
        & LHC / $\tau$-decays~\cite{Cirigliano:2021img}\\
        $[\epsilon_P^{cs}]_{\tau\mu}$  
        & {$\bf2.0 \times 10^{-1}$}  % AF and JK
        &$\bf{\Gamma_{D_s \to \tau \nu}}$ 
        & \num{2.5e-2} % 
        & LHC~\cite{Fuentes-Martin:2020lea}\\
        $[\epsilon_P^{cs}]_{\tau\tau}$ 
        & {$\bf3.2 \times 10^{-2}$ } 
        & $\bf{\Gamma_{D_s \to \tau \nu}}$ 
        & \num{2.5e-2} 
        & LHC~\cite{Fuentes-Martin:2020lea}\\
         \bottomrule
    \end{tabular}
    }
    \caption{Summary of 90\,\% CL bounds on the absolute value of pseudoscalar WEFT Wilson coefficients. The results in the left column are independent of the underlying high-energy theory, whereas the right column shows bounds derived under the assumption that SMEFT is the underlying theory at energies above the electroweak scale. Empty cells in the right column indicate operators for which we are unaware of a SMEFT bound in the literature, or for which the SMEFT bound is significantly weaker than the WEFT constraint. Bounds shown in bold face have been calculated in this work. Bounds marked with $(*)$ are given at the 95\,\% rather than 90\,\% CL. 
    }
    \label{tab:external-constraints-P}
\end{table}
%=======================================================================

%=======================================================================
\begin{table}[t]
    \centering
    \scalebox{0.8}{
    \begin{tabular}{l|cc|cc}
        \toprule
        Coupling 
        &  \multicolumn{2}{c|}{Low energy (WEFT)} 
        & \multicolumn{2}{c}{High energy / CLFV (SMEFT)} \\
        & 90\,\% CL bound & process & 90\,\% CL bound & process\\
        \midrule
        $[\epsilon_R^{ud}]_{ee}$       
        & \bfseries\num{2.5e-3}  % MGA and JK
        & $\bf\Gamma_{\pi\to e\nu} / \Gamma_{\pi\to\mu\nu}$ 
        &
        &\\
        $[\epsilon_R^{ud}]_{e\mu}$     
        & \num{3.7e-2}  % MGA and JK
        &  $\nu_\mu\to\nu_e$~\cite{Astier:2001yj,Astier:2003gs,Biggio:2009nt}
        & -
        & - \\
        $[\epsilon_R^{ud}]_{e\tau}$    
        & \bfseries\num{4.1e-2} % MGA and JK  
        & $\bf\Gamma_{\pi\to e\nu} / \Gamma_{\pi\to\mu\nu}$ 
        & - 
        & - \\
        $[\epsilon_R^{ud}]_{\mu e}$  
        & \bfseries\num{7.1e-2}  % MGA and JK}
        & $\bf\Gamma_{\pi\to e\nu} / \Gamma_{\pi\to\mu\nu}$ 
        & - 
        & - \\
        $[\epsilon_R^{ud}]_{\mu\mu}$
        & \bfseries\num{2.5e-3}  % MGA and JK
        & $\bf\Gamma_{\pi\to e\nu} / \Gamma_{\pi\to\mu\nu}$
        &
        &\\
        $[\epsilon_R^{ud}]_{\mu\tau}$  
        & \bfseries\num{7.1e-2}  % MGA and JK}
        & $\bf\Gamma_{\pi\to e\nu} / \Gamma_{\pi\to\mu\nu}$ 
        & - 
        & - \\

        $[\epsilon_R^{ud}]_{\tau e}$   
        & \num{0.12}  % MGA and JK
        & $\nu_e\to\nu_\tau$~\cite{Astier:2001yj,Astier:2003gs,Biggio:2009nt}
        & - 
        & - \\
        $[\epsilon_R^{ud}]_{\tau\mu}$
        & \num{1.8e-2} % MGA and JK
        & $\nu_\mu\to\nu_\tau$~\cite{Astier:2001yj,Astier:2003gs,Biggio:2009nt}
        & - 
        & - \\
        $[\epsilon_R^{ud}]_{\tau\tau}$ & \num{6.1e-3}   
        & $\tau$-decay~\cite{Cirigliano:2018dyk}  
        &
        &\\
        \midrule
        %------------------------------------------------------------
        $[\epsilon_R^{us}]_{ee}$       
        & \bfseries\num{5.2e-3}  % MGA and JK  
        & $\bf\Gamma_{K\to e\nu} / \Gamma_{K\to\mu\nu}$ 
        &
        &\\
        $[\epsilon_R^{us}]_{e\mu}$    
        & \bfseries\num{1.0e-1}  % MGA and JK  
        & $\bf\Gamma_{K\to e\nu} / \Gamma_{K\to\mu\nu}$
        & - 
        & - \\
        $[\epsilon_R^{us}]_{e\tau}$    
        & \bfseries\num{1.0e-1}  % MGA and JK  
        & $\bf\Gamma_{K\to e\nu} / \Gamma_{K\to\mu\nu}$
        & - 
        & - \\
        $[\epsilon_R^{us}]_{\mu e}$  
        & \bfseries\num{5.5e-2}  % MGA and JK  
        & $\bf\Gamma_{K\to e\nu} / \Gamma_{K\to\mu\nu}$ 
        & - 
        & - \\
        $[\epsilon_R^{us}]_{\mu\mu}$   
        & \bfseries\num{5.2e-3} % MGA and JK  
        & $\bf\Gamma_{K\to e\nu} / \Gamma_{K\to\mu\nu}$
        &
        &\\
        $[\epsilon_R^{us}]_{\mu\tau}$  
        & \bfseries\num{5.5e-2}  % MGA and JK  
        & $\bf\Gamma_{K\to e\nu} / \Gamma_{K\to\mu\nu}$ 
        & - 
        & - \\
        $[\epsilon_R^{us}]_{\tau e}$  
        &   \bfseries\num{10.7e-2}
        & $\bf\Gamma_{\tau \to K \nu}/\Gamma_{K \to \mu \nu}$ 
        & - 
        & - \\
        $[\epsilon_R^{us}]_{\tau\mu}$  
        &   \bfseries\num{10.7e-2}
        & $\bf\Gamma_{\tau \to K \nu}/\Gamma_{K \to \mu \nu}$ 
        & - 
        & - \\
        $[\epsilon_R^{us}]_{\tau\tau}$ 
        & \num{1.5e-2}  % MGA, AF and JK
        & $\tau$-decays~\cite{Gonzalez-Solis:2019owk}
        &
        &\\
        \midrule
        %------------------------------------------------------------
        $[\epsilon_R^{cs}]_{ee}$       
        & \bfseries\num{1.3e-2}%$\bf1.3 \times 10^{-2}$ % AF and JK
        & $\bf\Gamma_{D\to K\mu \nu}/\Gamma_{D\to Ke\nu}$
        &
        &\\
        $[\epsilon_R^{cs}]_{e\mu}$    
        & \bfseries\num{0.13}  % AF and JK
        & $\bf\Gamma_{D\to K\mu \nu}/\Gamma_{D\to Ke\nu}$
        & - 
        & - \\
        $[\epsilon_R^{cs}]_{e\tau}$    
        & \bfseries\num{0.13}  % AF and JK
        & $\bf\Gamma_{D\to K\mu \nu}/\Gamma_{D\to Ke\nu}$
        & - 
        & - \\
         $[\epsilon_R^{cs}]_{\mu e}$  
        & \bfseries\num{0.16} % AF and JK
        &  $\bf\Gamma_{D\to K\mu \nu}/\Gamma_{D\to Ke\nu}$ 
        & - 
        & - \\
        $[\epsilon_R^{cs}]_{\mu\mu}$   
        &  \bfseries\num{1.3e-2}  % AF and JK
        & $\bf\Gamma_{D\to K\mu \nu}/\Gamma_{D\to Ke\nu}$ 
        &
        &\\
        $[\epsilon_R^{cs}]_{\mu\tau}$  
        & \bfseries\num{0.16} % AF and JK
        &  $\bf\Gamma_{D\to K\mu \nu}/\Gamma_{D\to Ke\nu}$ 
        & - 
        & - \\
        $[\epsilon_R^{cs}]_{\tau e}$  
        &   \bfseries\num{3.2e-1} % AF and JK
        &  $\bf\Gamma_{D_s \to \tau \nu}$ 
        & - 
        & - \\
        $[\epsilon_R^{cs}]_{\tau\mu}$  
        &  \bfseries\num{3.2e-1} % AF and JK
        &  $\bf\Gamma_{D_s \to \tau \nu}$ 
        & - 
        & - \\
        $[\epsilon_R^{cs}]_{\tau\tau}$
        &  \bfseries\num{5.0e-2} % AF and JK  
        &  $\bf\Gamma_{D_s \to \tau \nu}$
        &
        &\\
        \bottomrule
    \end{tabular}}
    \caption{Summary of 90\,\% CL bounds on the absolute value of right-handed WEFT Wilson coefficients. The results in the left column are independent of the underlying high-energy theory. In the right-hand column we indicate with dashes the lepton flavor off-diagonal operators, which are not generated from dimension-6 SMEFT operators. Empty cells in the right column indicate operators for which we are unaware of a SMEFT bound in the literature, or for which the SMEFT bound is significantly weaker than the WEFT constraint. Bounds shown in bold face have been calculated in this work.
    }
    \label{tab:external-constraints-R}
\end{table}
%=======================================================================

%=======================================================================
\begin{table}[t]
    \centering
    \scalebox{0.8}{
    \begin{tabular}{l|cc|cc}
        \toprule
        Coupling 
        &  \multicolumn{2}{c|}{Low energy (WEFT)} 
        & \multicolumn{2}{c}{High energy / CLFV (SMEFT)} \\
        & 90\,\% CL bound 
        & process & 90\,\% CL bound & process\\
        \midrule
        $[\epsilon_S^{us}]_{ee}$       
        & \num{7.0e-3} 
        & $K$-decays~\cite{Gonzalez-Alonso:2016etj} % eq. 50
        &
        &\\
        $[\epsilon_S^{us}]_{e\mu}$    
        & $\bf{\cal O}(10^{-2})$ %\num{7.0e-3} \jk{(*)} 
        & $\bf K$-{\bf decays} % eq. 50 
        & \bf{\num{6.2e-7}}
        & \bf{$\mu \to e$ conversion} \\
        $[\epsilon_S^{us}]_{e\tau}$    
        & $\bf{\cal O}(10^{-2})$ %\num{7.0e-3} \jk{(*)}
        & $\bf K$-{\bf decays} % eq. 50
        & \num{7.1e-2}
        & LHC~\cite{Cirigliano:2021img}\\
        $[\epsilon_S^{us}]_{\mu e}$  
        & $\bf{\cal O}(10^{-2})$ 
        & $\bf K$-{\bf decays}
        & \bf{\num{6.2e-7}}
        & \bf{$\mu \to e$ conversion} \\
        $[\epsilon_S^{us}]_{\mu\mu}$   
        & \num{1.3e-3}  % AF and JK
        & $K$-decays~\cite{Gonzalez-Alonso:2016etj} % eq. 45
        &
        &\\
        $[\epsilon_S^{us}]_{\mu\tau}$  
        & $\bf{\cal O}(10^{-2})$ 
        & $\bf K$-{\bf decays}
        &
        &\\
        $[\epsilon_S^{us}]_{\tau e}$  
        & $\bf{\cal O}(10^{-1})$
        &  $\bf {\tau}$~{\bf decays} 
        & \bfseries{\num{3.1e-2}${}^{(*)}$} % JK and AF
        & {\bf LHC (data~\cite{Aaboud:2018vgh})}\\
        $[\epsilon_S^{us}]_{\tau\mu}$  
        & $\bf{\cal O}(10^{-1})$
        &  $\bf {\tau}$~{\bf decays} 
        & \bfseries{\num{3.1e-2}${}^{(*)}$} % JK and AF 
        & {\bf LHC (data~\cite{Aaboud:2018vgh})}\\
        $[\epsilon_S^{us}]_{\tau\tau}$ 
        & \num{2.4e-2}  % MGA, AF and JK
        & $\tau$ decays ~\cite{Gonzalez-Solis:2019owk}
        & \bfseries{\num{3.1e-2}${}^{(*)}$} % JK and AF
        & {\bf LHC (data~\cite{Aaboud:2018vgh})}\\
       
        %--------------------------------------------------------------------
        \midrule\midrule
        %--------------------------------------------------------------------
        $[\epsilon_T^{ud}]_{ee}$       
        & \num{2.1e-3}  
        & $\beta$-decays~\cite{Falkowski:2020pma}
        &
        &\\
        $[\epsilon_T^{ud}]_{e\mu}$    
        & \bfseries\num{3.3e-2}
        & $\bf\beta$-{\bf decays}
        & \bf{\num{3.6e-7}}
        & \bf{$\mu \to e$ conversion} \\
        $[\epsilon_T^{ud}]_{e\tau}$    
        & \bfseries\num{3.3e-2}
        & $\bf\beta$-{\bf decays}
        &\num{1.1e-3}/\num{5.7e-4}
        & LHC~\cite{Cirigliano:2021img}/$\tau$ decays~\cite{Cirigliano:2021img}\\
         $[\epsilon_T^{ud}]_{\mu e}$  
        & -
        &
        & \bf{\num{3.6e-7}}
        & \bf{$\mu \to e$ conversion} \\
        $[\epsilon_T^{ud}]_{\mu\mu}$   
        & -  
        &  
        &
        &\\
        $[\epsilon_T^{ud}]_{\mu\tau}$  
        & -
        &
        & $\mathcal{O}(\num{e-4})$
        & LHC top decays~\cite{Cirigliano:2021img,ATLAS:2018avw} \\
        $[\epsilon_T^{ud}]_{\tau e}$  
        & $\bf{\cal O}(10^{-1})$
        & $\bf\tau$~{\bf{decays}} 
        & \num{1.1e-3}/\num{5.7e-4}
        & LHC~\cite{Cirigliano:2021img}/$\tau$ decays~\cite{Cirigliano:2021img}\\
        $[\epsilon_T^{ud}]_{\tau\mu}$  
        & $\bf{\cal O}(10^{-1})$
        & $\bf\tau$~{\bf{decays}} 
        & \num{5.2e-3}${}^{(*)}$ % AF and YS
        & LHC~\cite{Cirigliano:2018dyk}\\
        $[\epsilon_T^{ud}]_{\tau\tau}$ 
        & \num{1.3e-2}  
        & $\tau$-decays~\cite{Cirigliano:2018dyk}
        & \num{5.2e-3}${}^{(*)}$ % AF and YS
        & LHC~\cite{Cirigliano:2018dyk}\\
        %--------------------------------------------------------------------
        \midrule
        %--------------------------------------------------------------------
        $[\epsilon_T^{cs}]_{ee}$       
        &\bfseries\num{0.62}  % AF and JK
        & $\bf\Gamma_{D \to K \mu \nu}/\Gamma_{D \to K e \nu}$ 
        & \num{8.8e-3}
        & LHC~\cite{Fuentes-Martin:2020lea}\\
        $[\epsilon_T^{cs}]_{e\mu}$    
        & \bfseries\num{0.61} % AF and JK
        & $\bf\Gamma_{D \to K \mu \nu}/\Gamma_{D \to K e \nu}$ 
        & \num{8.8e-3}
        & LHC~\cite{Fuentes-Martin:2020lea}\\
        $[\epsilon_T^{cs}]_{e\tau}$    
        & \bfseries\num{0.61}  % AF and JK
        & $\bf\Gamma_{D \to K \mu \nu}/\Gamma_{D \to K e \nu}$ 
        & \num{8.8e-3}
        / \num{1.3e-4}
        & LHC~\cite{Fuentes-Martin:2020lea}/$\tau$ decays~\cite{Cirigliano:2021img}\\
        $[\epsilon_T^{cs}]_{\mu e}$  
        & \bfseries\num{0.76}  % AF and JK
        & $\bf\Gamma_{D \to K \mu \nu}/\Gamma_{D \to K e \nu}$ 
        &  \num{1.2e-2} 
        & LHC~\cite{Fuentes-Martin:2020lea}\\
        $[\epsilon_T^{cs}]_{\mu\mu}$   
        & \bfseries\num{0.21}  % AF and JK
        & $\bf\Gamma_{D \to K \mu \nu}/\Gamma_{D \to K e \nu}$ 
        &  \num{1.2e-2} 
        & LHC~\cite{Fuentes-Martin:2020lea}\\
        $[\epsilon_T^{cs}]_{\mu\tau}$  
        & \bfseries\num{0.76}  % AF and JK
        & $\bf\Gamma_{D \to K \mu \nu}/\Gamma_{D \to K e \nu}$ 
        &  \num{1.2e-2} 
        & LHC~\cite{Fuentes-Martin:2020lea}\\
        $[\epsilon_T^{cs}]_{\tau e}$  
        & -
        & 
        &\num{1.1e-2} / \num{1.3e-4}

        & LHC~\cite{Cirigliano:2021img}/ $\tau$ decay~\cite{Cirigliano:2021img}\\
        $[\epsilon_T^{cs}]_{\tau\mu}$  
        & - 
        &
        & \num{1.6e-2}  
        & LHC~\cite{Fuentes-Martin:2020lea}\\
        $[\epsilon_T^{cs}]_{\tau\tau}$ 
        & - 
        &
        & \num{1.6e-2}  
        & LHC~\cite{Fuentes-Martin:2020lea}\\
         \bottomrule
    \end{tabular}
    }
    \caption{Summary of 90\,\% CL bounds on the absolute value of scalar and tensor WEFT Wilson coefficients. The results in the left column are independent of the underlying high-energy theory, whereas the right column shows bounds derived under the assumption that SMEFT is the underlying theory at energies above the electroweak scale. Empty cells in the right column indicate operators for which we are unaware of a SMEFT bound in the literature, or for which the SMEFT bound is significantly weaker than the WEFT constraint. Bounds shown in bold face have been calculated in this work. Bounds marked with $(*)$ are given at the 95\,\% rather than 90\,\% CL.
    }
    \label{tab:external-constraints-ST}
\end{table}
%=======================================================================

%-----------------------------------------------------------------------
\subsection{Neutrino experiments}
%-----------------------------------------------------------------------

From a qualitative point of view, the bounds from the NOMAD experiment~\cite{Astier:2001yj, Astier:2003gs} are equivalent to those expected from FASER$\nu$ in the sense that they depend on the same physical processes.  
NOMAD bounds on $[\epsilon_R^{ud}]_{e\mu,\tau e,\tau\mu}$ have been evaluated in~\cite{Biggio:2009nt} and have been found to be at the $\cO(10^{-2}-10^{-3})$ level.

For scalar and tensor interactions, the only limits from neutrino experiments available in the literature are from reactor data, $\bar{\nu}_e \to \bar{\nu}_e$~\cite{Falkowski:2019xoe}.  
The relevant observables are sensitive to the $[\epsilon^{ud}_X]_{e\mu}$ and $[\epsilon^{ud}_X]_{e\tau}$ coefficients with $X=R,S,T$ at the linear order. 
The current bounds on $[\epsilon^{ud}_X]_{e\mu}$ and $[\epsilon^{ud}_X]_{e\tau}$ are in the $0.05$--$0.10$ range~\cite{Falkowski:2019xoe}.

We are not aware of limits on pseudoscalar interactions from neutrino experiments.

%-----------------------------------------------------------------------
\subsection{(Semi-)leptonic Hadron Decays and $\beta$-decays}
%-----------------------------------------------------------------------

Precision measurements of leptonic and semi-leptonic hadron decay rates as well as $\beta$-decay rates are sensitive to the lepton-flavor off-diagonal EFT coefficients at quadratic order, similarly to FASER$\nu$. 
Compared to FASER$\nu$, meson decay measurements typically benefit from higher statistics because they do not require neutrino detection. 
This leads to a typical uncertainty in the $10^{-2}$--$10^{-3}$ range for the event rates, and consequently in the $10^{-1}$--$10^{-2}$ range for the EFT coefficients, except when additional enhancements are present, such as for pseudoscalar operators.  Limits on flavor-diagonal coefficients are typically stronger than those on off-diagonal coefficients because the corresponding amplitudes can interfere with the SM ones, leading to sensitivity already at the linear order~\cite{Gonzalez-Alonso:2016etj, Cirigliano:2018dyk, Falkowski:2020pma}.

We can classify the processes sensitive to the $[\epsilon^{ud}_X]_{\alpha\beta}$ coefficients in several groups:
\begin{itemize}
    \item 
    {\bf $\beta$-decays} are sensitive to the $ee$, $e\mu$ and $e\tau$ elements of the $[\epsilon^{ud}_X]_{\alpha\beta}$ matrices. 
    We follow the analysis of ref.~\cite{Falkowski:2020pma} modified so as to include the effects from lepton-flavor off-diagonal Wilson coefficients. We find that the  $[\epsilon^{ud}_T]_{e \mu}$ and $[\epsilon^{ud}_T]_{e \tau}$  are bound to be smaller than $\num{3e-2}$, while the constraint on the $ee$ coefficient is about an order of magnitude stronger. 
 
    \item
    {\bf Leptonic pion decays} ($\pi\to e\nu_{\mu,\tau}, \mu\nu_{e,\tau}$) are sensitive to the $e\mu$, $e\tau$, $\mu e$ and $\mu\tau$ coefficients of the axial and pseudoscalar interactions, i.e.\ $\epsilon^{ud}_{R}$ and $\epsilon^{ud}_{P}$.
    The pseudoscalar contribution enjoys large chiral enhancement, see \cref{sec:production}, which translates into strong bounds. 
    Using the ratio $\Gamma(\pi\to e\nu) / \Gamma(\pi\to\mu\nu)$~\cite{Zyla:2020zbs, Cirigliano:2007xi} and switching on one operator at a time one obtains bounds at $\cO(10^{-6})$ and $\cO(10^{-3})$ for lepton-flavor off-diagonal couplings to electrons and muons, respectively \cite{Falkowski:2019xoe}, and at the $10^{-7}$ and $10^{-4}$ level for the diagonal $ee$ and $\mu\mu$ couplings, respectively~\cite{Gonzalez-Alonso:2016etj}. 
    Note that, if couplings to both electrons and muons were present, and if the corresponding Wilson coefficients scale with the charged lepton masses, the two contributions would cancel in the ratio $\Gamma(\pi\to e\nu) / \Gamma (\pi\to\mu\nu)$, see e.g.~\cite{Bhattacharya:2011qm}. 
    In this case, slightly weaker bounds can still be obtained from the individual decay widths using the decay constants calculated on the lattice~\cite{Gonzalez-Alonso:2016etj,Aoki:2019cca}. 
    Even these constraints can be relaxed if one allows for fine-tuned cancellations between diagonal and off-diagonal terms~\cite{Bhattacharya:2011qm}. 
    Indirectly these processes also probe scalar and tensor couplings as these mix with the pseudoscalar couplings through loops~\cite{Gonzalez-Alonso:2017iyc, Falkowski:2019xoe}. 
    
    \item 
    {\bf Hadronic $\tau$ decays} are sensitive to the $\tau e$, $\tau \mu$, and $\tau\tau$ elements of the $\epsilon_X^{ud}$ matrices, with any Lorentz structure. Lepton-flavor diagonal coefficients were studied in ref.~\cite{Cirigliano:2018dyk}, leading to constraints in the $10^{-2}$--$10^{-3}$ range. 
    Since decay rates are sensitive to the off-diagonal Wilson coefficients only at quadratic order, the bounds on these coefficients are expected to be weaker, in the $10^{-1}$--$10^{-2}$ range.
    We derived such bounds for the pseudoscalar $[\epsilon_P^{ud}]_{\tau e,\tau\mu}$ coefficients from $\tau \to \pi \nu$ decays. Deriving precise bounds for the scalar and tensor coefficients would require a nontrivial dedicated analysis, therefore we only give an order-of-magnitude estimate in 
    \cref{tab:external-constraints-ST}. 
    
\end{itemize}
Similar to the bounds on $ud$ couplings extracted from the rates of processes involving pions and neutrons, one can also derive bounds on the $us$ coefficients by replacing pions and neutrons by kaons and hyperons. 
For instance, bounds at the level of $10^{-5}$ can be obtained for the pseudoscalar coefficients $[\epsilon_P^{us}]_{e\mu,e\tau}$ from the ratio of $K \to \mu \nu$ and  $K \to e \nu$ decays. 
Strong bounds on the $us$ coefficients can be obtained as well from semileptonic kaon decays. 
Typical analyses of strange decays have focused on lepton-flavor diagonal coefficients, e.g.~\cite{Gonzalez-Alonso:2016etj}, which interfere with the SM contribution. 
Bounds on $[\epsilon^{us}_{S}]_{ee}$ can be derived based on the effect of these couplings on the kinematics of $K \to \pi e \nu$ decays parameterized in terms of the Dalitz plot~\cite{Gonzalez-Alonso:2016etj}. 
Since the dependence on the Wilson coefficients is mainly quadratic, we estimate similar ${\cal O} (10^{-2})$ bounds on the $e\mu$ and $e\tau$ couplings. 
 Comparable bounds can also be expected on $\mu e$ and $\mu\tau$ coefficients using $K \to \pi\mu\nu$ data, although such an analysis has never been carried out as far as we know. Finally, $[\epsilon^{us}_S]_{\mu\mu}$ was constrained in Ref.~\cite{Gonzalez-Alonso:2016etj} using $K \to \pi \mu \nu$ data and lattice input on the scalar form factor.

Limits on the $cs$ coefficients can be obtained from (semi)leptonic $D$ meson decays, see e.g.\  ref.~\cite{Becirevic:2020rzi} for a recent such analysis. 
We derive these limits by comparing the theoretically predicted event rates with the new physics corrections discussed in \cref{sec:Formalism} to the experimentally observed ones. The latter are taken from the 2020 edition of the Particle Data Group review, ref.~\cite{Zyla:2020zbs}, while for theoretical input such as form factors from the lattice, we rely on the FLAG 2019 Review~\cite{Aoki:2019cca}. Note that the tauonic coefficients $[\epsilon_T^{cs}]_{\tau\alpha}$ are unconstrained because (i)~they do not contribute to $D_s\to\tau\nu$ and (ii)~the semileptonic decay $D \to K \tau \nu$ is kinematically forbidden.

%-----------------------------------------------------------------------
\subsection{Collider bounds}
%-----------------------------------------------------------------------

If the SMEFT is a valid theory at LHC scales, then WEFT coefficients can be written as a linear combination of SMEFT coefficients (``matching")~\cite{Jenkins:2017jig,Cirigliano:2012ab}. 
We use tree-level matching to translate collider bounds on SMEFT operators into constraints on WEFT operators (see ref.~\cite{Dekens:2019ept} for loop corrections).
For the running and mixing effects between the operators we follow~ref.~\cite{Gonzalez-Alonso:2017iyc}.
Besides requiring the validity of these approximations, we also assume that contributions from higher-dimensional SMEFT operators ($\text{dim} > 6$) are negligible. 
This is not a trivial assumption because LHC bounds are often dominated by processes that depend quadratically on the dimension-6 coefficients.

For (pseudo-)scalar and tensor operators ($X=P,S,T$), which flip chirality, interference with chirality-conserving SM processes is effectively zero at high energies. 
As a result, collider observables are only quadratically sensitive to $\epsilon_{P,S,T}$, and thus the bounds that have been obtained on lepton flavor-diagonal operators $[\epsilon_{P,S,T}^{jk}]_{\alpha\alpha}$ should actually be interpreted as bounds on the incoherent sum over all three neutrino flavors, $\sum_\beta [\epsilon_{P,S,T}^{jk}]_{\alpha\beta}$. 
This allows us to write the bounds obtained from $pp \to  e + \text{MET}$ (ATLAS, \SI{13}{TeV}, \SI{36}{fb^{-1}}~\cite{Aaboud:2017efa}) in ref.~\cite{Gupta:2018qil} as $[\epsilon^{ud}_{X}]_{e\alpha}<\num{2e-3}$ for $\alpha=e,\mu,\tau$ and $X=S,P,T$.
The same dataset~\cite{Aaboud:2017efa} can be used to search for operators where the down quark is replaced by a strange quark, although with less sensitivity due to PDF suppression. 
Bounds on couplings involving muons are obtained in the same way as those on couplings involving electrons, while constraints on couplings to $\tau$ leptons are slightly weaker~\cite{Gonzalez-Alonso:2016etj,Falkowski:2017pss,Cirigliano:2018dyk}.
The bounds on the $[\epsilon^{ud}_X]_{\tau\alpha}$ ($X=S,P,T$) based on $pp\to \tau + \text{MET}$ (ATLAS, 13\,TeV, 36\,fb$^{-1}$~\cite{Aaboud:2018vgh}) were evaluated in~\cite{Cirigliano:2018dyk} and found to be $\cO(10^{-3})$. 

Here we perform a similar analysis to ref.~\cite{Cirigliano:2018dyk} in order to put upper bounds on $[\epsilon^{us}_X]_{\tau\alpha}$ ($X=S,P$).  We simulate the new physics signal ($\tau+\text{MET}$) in \textsc{MadGraph~5} v\,2.9.2~\cite{Alwall:2014hca} with the SMEFTsim plugin~\cite{Brivio:2017btx,Brivio:2020onw} for the hard process, \textsc{Pythia}~v\,8.2~\cite{Sjostrand:2014zea} for the showering and hadronization, and \textsc{Delphes}~v\,3.4.2~\cite{deFavereau:2013fsa} as a simple detector simulation. In Delphes, we use the default implementation of the ATLAS detector, but we set the tau tagging efficiency to one in order to prevent Delphes from discarding events. Instead, we later apply weight factors corresponding to the tau tagging efficiencies reported in ref.~\cite{Aaboud:2018vgh}. That way, we make optimal use of the generated Monte Carlo statistics (\SI{10000}{events} for each of the considered SMEFT couplings $[c_{\ell e d Q}]_{3321}$, $[c_{\ell e Q u}^{(1)}]_{3321}$, $[c_{\ell e Q u}^{(3)}]_{3321}$) and avoid discarding events. Following ref.~\cite{Aaboud:2018vgh}, we take the tau tagging efficiency to be 60\% at transverse momentum $p_{T,\tau} < \SI{100}{GeV}$, 30\% at $p_{T,\tau} > \SI{2}{TeV}$, and we interpolate linearly in between. In a similar way, we also apply trigger efficiencies, which are 98\% at missing transverse momentum $\slashed{p}_T > \SI{250}{GeV}$ and 80\% at $\slashed{p}_T = \SI{150}{GeV}$; we once again use linear interpolation in between. We apply the following cuts: there must be at least one $\tau$ lepton in the event, the missing transverse momentum must by larger than \SI{150}{GeV}, the leading $\tau$ must satisfy $p_{T,\tau} > \SI{50}{GeV}$ and $|\eta_\tau| < 2.4$ ($\tau$s in the pseudorapidity window $1.37 < |\eta_\tau| < 1.52$ are excluded, though), the transverse momentum ratio $p_{T,\tau} / \slashed{p}_T$ must be between 0.7 and 1.3, and the azimuthal angle between the $\tau$ and the missing momentum vector must be $\Delta\phi(\tau, \slashed{p}_T) > 2.4$.  We have verified that our simulation reproduces the SM transverse mass distribution shown in fig.~1 of ref.~\cite{Aaboud:2018vgh}, especially its high-$m_T$ tail, very well when all new physics effects are switched off. We can therefore confidently use it to set limits on the aforementioned SMEFT operators. We do so for one operator at a time by determining the value of its Wilson coefficient at which the predicted signal cross-section matches the 95\,\% confidence level bound shown in fig.~2 of ref.~\cite{Aaboud:2018vgh}. We do this for different $m_T$ thresholds, and we pick the one that gives the optimal limit. We find that upper bounds on the Wilson coefficients are $\cO(\text{few} \times 10^{-3})$.  Once again, we perform the matching between SMEFT and WEFT, see for instance ref.~\cite{Falkowski:2019xoe}. We finally run the Wilson coefficients down to the \SI{2}{GeV} scale using the relations given in ref.~\cite{Gonzalez-Alonso:2017iyc}. For WEFT operators that can originate from more than one SMEFT operator we use the limit obtained when switching on the most weakly constrained SMEFT operator only. Our final result -- a limit of $[\epsilon^{us}_{S,P}]_{\tau\alpha} < \num{5.8e-3}$ -- is quoted in \cref{tab:external-constraints-P,tab:external-constraints-ST}.

For operators with couplings to charm and strange quarks, finally, we adopt the LHC bounds from ref.~\cite{Fuentes-Martin:2020lea}.

%-----------------------------------------------------------------------
\subsection{Charged-lepton flavor violation}
%-----------------------------------------------------------------------

Assuming SMEFT is the UV-completion of WEFT, the same dimension-6 operators that generate lepton-flavor off-diagonal $\epsilon^{jk}_X$ coefficients in the low-energy (WEFT) limit also generate neutral current interactions between quarks and two charged leptons of different flavor. 
Such charged-lepton-flavor violating~(CLFV) interactions are strongly constrained because they generate processes that are forbidden in the SM, such as $\mu \to e\gamma$, $\tau \to \mu \gamma$, or $\mu \to e$ conversion on nuclei. Here, we focus on tree-level effects of this type. Following the discussion in ref.~\cite{Cirigliano:2009bz}, the experimental bounds on  $\mu \to e$ conversion on gold nuclei~\cite{Bertl:2006up} constrain the SMEFT Wilson coefficients $[c_{lequ}^{(1)}]_{e \mu 11}$ and $[c_{ledq}]_{e \mu 11}$  to be smaller than $\cO((100~\tev)^{-2})$, assuming only one SMEFT operator present at a time. Also the tensor operator $[c_{lequ}^{(3)}]_{e \mu 11}$ can be constrained thanks to its renormalization group mixing with $[c_{lequ}^{(1)}]_{e \mu 11}$. The resulting limit is about an order of magnitude weaker than the ones on $[c_{lequ}^{(1)}]_{e \mu 11}$ and $[c_{ledq}]_{e \mu 11}$. In our calculation, we include RGE running according to ref.~\cite{Cirigliano:2017azj}, and we use the wave function overlap integrals from ref.~\cite{Kitano:2002mt}.  As the SMEFT operators $[c_{lequ}^{(1)}]_{e \mu 11}$, $[c_{ledq}]_{e \mu 11}$, and $[c_{lequ}^{(3)}]_{e \mu 11}$ contribute to charged current interactions as well, the strong limits that $\mu \to e$ conversion imposes on them translate into equally strong bounds on their WEFT counterparts. In particular, we find $[\epsilon_{S,P}^{ud}]_{\mu e} \lesssim \num{2e-8}$ and $[\epsilon_T^{ud}]_{\mu e} \lesssim \num{4e-7}$.  The same bounds hold for $[\epsilon_{S,P}^{ud}]_{e \mu}$ and $[\epsilon_T^{ud}]_{e\mu}$, while slightly weaker bounds are found for operators involving the strange quark: $[\epsilon_{S,P}^{us}]_{\mu e,e\mu} \lesssim \num{6e-7}$ and $[\epsilon_T^{us}]_{e\mu,\mu e} \lesssim \num{7e-6}$.

Constraints on $\tau \to e$ transitions were recently discussed in ref.~\cite{Cirigliano:2021img}, 
who found $\cO(10^{-4})$ constraints on the tensor and pseudoscalar coefficients $[\epsilon_{T,P}^{ud}]_{\tau e, e \tau}$.

We stress again that both collider and CLFV bounds do not hold when SMEFT is not a valid effective theory above the electroweak scale, for instance because new particles exist at or below the electroweak scale.

%=======================================================================
\section{Discussions and Conclusions}
\label{sec:conclusion}
%=======================================================================

In summary, we have highlighted the significant potential of the FASER$\nu$ detector at CERN to constrain new physics affecting neutrino interactions with matter.  
Working in Weak Effective Field Theory (the most general effective theory at $\mu \sim$~few~GeV, assuming the absence of non-SM particles at this scale), we have identified the charged-current dimension-6 operators that modify the observable neutrino rates in FASER$\nu$.
The effects of the corresponding Wilson coefficients $[\epsilon_X^{jk}]_{\alpha\beta}$  are encapsulated in  the modified neutrino oscillation ``probability'' $\tilde{P}_{\alpha\beta}$. 
In this object, the Wilson coefficients are weighted by so-called production and detection coefficients $p_{XY,\alpha}^{S,jk}$ and $d_{XY,\alpha}^{jk}$, which  describe how neutrino production and detection rates  differ from their counterparts in the SM.  
We evaluated the production and detection coefficients  for the dimension-6 operators with all possible Lorentz structures $X, Y = L, R, S, P, T$.
This amounts to a comprehensive characterization of heavy new physics in FASER$\nu$ at the leading order in the relevant EFT.  

We have found that neutrino production in fully leptonic meson decays, which accounts for most of FASER$\nu$'s total neutrino flux, enjoys a particularly strong enhancement if there are new pseudoscalar interactions. 
The reason is that such interactions do not suffer from chiral suppression, unlike the SM $V$--$A$ interactions. 
Consequently, the Wilson coefficients of some pseudoscalar operators can be constrained at the per mille level ($[\epsilon_P^{jk}]_{\alpha\beta} \lesssim \num{e-3}$), corresponding to sensitivity to new physics scales up to $\sim \SI{10}{TeV}$. 
We also find good sensitivity to some operators that lead to an anomalous $\nu_\tau$ flux, or to the creation of $\tau$ leptons in charged-current interactions of $\nu_e$ and $\nu_\mu$. 
The reason for this is the very low $\nu_\tau$ background in the Standard Model. 
Most other interactions will be constrained down to $[\epsilon_X^{jk}]_{\alpha\beta} \sim \text{0.01--0.1}$, corresponding new physics at the TeV scale.

Compared to existing limits from other experiments, FASER$\nu$ will, for a number of operators, reach similar sensitivities, showing that LHC neutrinos offer an interesting new way of probing physics beyond the Standard Model. Unlike other probes (meson decays, ATLAS and CMS analyses, etc.) a collider neutrino experiment like FASER$\nu$ has the unique capability to identify the neutrino flavor. This is crucial complementary information in case excesses are found elsewhere in the future. Moreover, it allows to lift parameter degeneracies that may affect the interpretation of other measurements. One can, for instance, imagine a situation where different new physics effects with different signs conspire to leave a given meson or $\tau$ decay branching ratio unchanged compared to the SM, but change the flavor of the emitted neutrinos. Only a neutrino detector like FASER$\nu$ would be able to uncover such a conspiracy.

Our main results -- the FASER$\nu$ sensitivity estimates and their comparison to other constraints -- are summarized in \cref{fig:constraints-summary-P,fig:constraints-summary-R,fig:constraints-summary-S,fig:constraints-summary-T}.

We conclude that even a relatively cheap experiment like FASER$\nu$ can make important contributions to neutrino physics, indicating that neutrinos produced in LHC collisions offer interesting untapped potential for discovery. 
Exploiting this potential with FASER$\nu$, in the recently approved SND@LHC detector~\cite{Ahdida:2020evc}, and in other future projects will give a whole new dimension to the LHC physics program and thus benefit both the collider community and the neutrino community.

%=======================================================================
\begin{acknowledgments}
It is our great pleasure to thank Felix Kling for many useful discussions on the FASER experiment, for providing the neutrino fluxes in machine-readable form, and for invaluable comments on the manuscript.
Moreover, we are grateful to Sacha Davidson for valuable advice on the calculation of $\mu\to e$ conversion rates in nuclei.
Two babies were born during the completion of this project, and we thank them for their understanding. 
AF has received funding from the Agence Nationale de la Recherche (ANR) under grant ANR-19-CE31-0012 (project MORA) and from the European Union’s Horizon 2020 research and innovation programme under the Marie Skłodowska-Curie grant agreement No 860881-HIDDeN. 
MGA is supported by the {\it Generalitat Valenciana} (Spain) through the {\it plan GenT} program (CIDEGENT/2018/014). 
JK's work has been partially supported by the European Research Council (ERC)
under the European Union's Horizon 2020 research and innovation program (grant
agreement No.\ 637506, ``$\nu$Directions'').
YS is supported by the United States-Israel Binational Science Foundation (BSF) (NSF-BSF program grant No.~2018683), by the Israel Science Foundation (grant No.~482/20) and by the Azrieli foundation. 
YS is Taub fellow (supported by the Taub Family Foundation).  ZT is supported by the U.S. Department of Energy under the award number DE-SC0020250.
\end{acknowledgments}
%=======================================================================

%=======================================================================
\appendix
\section{Production coefficients in kaon decay}
\label{app:kaon}
%=======================================================================

In this appendix we provide details on the derivation of the production coefficients for kaon decays in \cref{sec:kaon-decays}.

%------------------------------------------------------------------
\subsection{Amplitudes}
%-------------------------------------------------------------------

We begin by writing down the amplitudes for the leptonic $K^\pm \to \ell^+_\alpha \nu$ and semileptonic  $K \to \pi^- \ell^+_\alpha \nu$ kaon decays mediated by the effective interactions in the Lagrangian of \cref{eq:EFT_lweft}. 

The case of leptonic decays is completely analogous to the pion decay discussed in \cref{sec:pion-decays}.
For the summed amplitude squared we can thus borrow \cref{eq:PION_sumAXY} and replace $\pi \to K$, $d \to s$:  
\bea
\label{eq:KAON_sumAXY-leptonic}
\sum | A_{L,\alpha}^{K,us}|^2  &= &  \frac{V_{us}^2  f_{K}^2   }{v^4} m_{\ell_\alpha}^2 \big (  m_{K}^2 - m_{\ell_\alpha}^2 \big ) , 
\nnl 
\sum A_{L,\alpha}^{K,us} \bar A_{P,\alpha}^{K,us}   &= & 
- \frac{V_{us}^2  f_{K}^2}{v^4} m_{\ell_\alpha} 
\big ( m_{K}^2 - m_{\ell_\alpha}^2 \big ) \frac{m_{K}^2}{m_u + m_s} , 
\nnl 
\sum | A_{P,\alpha}^{K,us}|^2  &= &   
\frac{V_{us}^2  f_{K}^2 }{v^4}    \big (  m_{K}^2 - m_{\ell_\alpha}^2 \big )  
\frac{m_K^4}{(m_u + m_s)^2 } . 
\eea

For the semileptonic decay the amplitude takes the form in \cref{eq:Mdecomposition} with the production amplitudes given by 
\begin{align}
A_{L,\alpha}^{K,us} = A_{R,\alpha}^{K,us}
        &= -\frac{V_{us}}{v^2} (\bar u_\nu \gamma_\mu P_L v_{\ell_\alpha})
            \bra{\pi^-} \bar{s} \gamma^\mu u \ket{K^0} \,,
        \label{eq:ALP-3b} \\
A_{S,\alpha}^{K,us}
        &= -\frac{V_{us}}{v^2} (\bar u_\nu P_R v_{\ell_\alpha})
            \bra{\pi^-} \bar{s} u \ket{K^0} \,,
        \label{eq:ASP-3b} \\
    A_{T,\alpha}^{K,us}
        &= -\frac{V_{us}}{2 v^2} (\bar u_\nu \sigma_{\mu\nu}P_R v_{\ell_\alpha})
            \bra{\pi^-} \bar{s} \sigma^{\mu\nu} u \ket{K^0} \,,
        \label{eq:ATP-3b} \\
    A_{P,\alpha}^{K,us} &= 0 \,, 
        \label{eq:APP-3b}
\end{align}
where  $v_{\ell_\alpha}$ and $\bar u_\nu$ are the spinor wave functions of the outgoing charged lepton and neutrino. 
In the tensor amplitude we have replaced 
$\bra{\pi^-} \bar{s} \sigma^{\mu\nu} P_R u \ket{K^0} \to \bra{\pi^-} \bar{s} \sigma^{\mu\nu} u \ket{K^0}$ 
using the Fierz identity $(\bar u \sigma_{\mu\nu}P_R v ) (\bar{s} \sigma^{\mu\nu} P_L u) =0$.
Moreover, we have used the fact that $\bra{\pi^-} \bar{s} \gamma^\mu \gamma^5 u \ket{K^0} = \bra{\pi^-} \bar{s} \gamma^5 u \ket{K^0} = 0$ due to parity conservation in QCD. 
For the non-zero hadronic matrix elements, we adopt the parametrization from~\cite{Antonelli:2008jg}:  
\begin{align}
    \braket{\pi^- | \bar{s} \gamma^\mu u | K^0}
        &= P^\mu f_+(q^2)+q^\mu f_-(q^2) \,,
        \label{eq:hadronic-ME-Kpi-V} \\[0.2cm]
    \braket{\pi^- | \bar{s} u | K^0}
        &= -\frac{m_K^2 - m_\pi^2}{m_s - m_u} f_0(q^2) \,,
        \label{eq:hadronic-ME-Kpi-S} \\
    \braket{\pi^- | \bar{s}\sigma^{\mu\nu} u | K^0}
        &= i \frac{p_K^\mu p_\pi^\nu - p_\pi^\mu p_K^\nu}{m_K} B_T(q^2) \,,
        \label{eq:hadronic-ME-Kpi-T}
\end{align}
where $P = p_\pi + p_K$ is the sum of the pion and kaon 4-momenta, $q = p_K - p_\pi=p_\nu + p_\ell$ is their difference. 
Using equations of motion one can derive the following relation between the three form factors $f_+$, $f_-$, and $f_0$: \begin{align}
    f_-(q^2) = \frac{m_K^2 - m_\pi^2}{q^2} \Big( f_0(q^2) - f_+(q^2) \Big) \,,
\end{align}
from which it also follows that $f_0(0) = f_+(0)$.
For the independent form factors $f_+(q^2)$, $f_0(q^2)$
we adopt the FlaviaNet dispersive parameterization~\cite{Antonelli:2010yf}:
\begin{align}
    f_+(q^2) &= f_+(0) + \Lambda_+ \frac{q^2}{m_{\pi}^2} + \cO(q^4) \,, \notag\\ 
    f_0(q^2) &= f_+(0) + \big( \log C - G(0) \big) \frac{m_\pi^2}{m_K^2 - m_\pi^2}
                                                   \frac{q^2}{m_{\pi}^2} + \cO(q^4) \,, 
\end{align}
where $G(0) = 0.0398(44)$ is calculated theoretically, 
and $\Lambda_+ = 0.02422(116)$ as well as $\log C = 0.1998(138)$ are obtained on the lattice~\cite{Carrasco:2016kpy}. The $N_f= 2 + 1 + 1$ value of $f_+(0)$ according to FLAG'19 is $f_+(0) = 0.9706(27)$~\cite{Aoki:2019cca}. 
For the tensor form factor we use the parameterization
\begin{align}
B_T(q^2) \approx B_T(0) \big ( 1 - s_T^{K\pi} q^2 \big ), 
\end{align} 
with $B_T(0)/f_+(0) = 0.68(3)$ and $s_T^{K\pi} = \SI{1.10(14)}{GeV^{-2}}$~\cite{Baum:2011rm}.

We plug the hadronic matrix elements from \cref{eq:hadronic-ME-Kpi-V,eq:hadronic-ME-Kpi-S,eq:hadronic-ME-Kpi-T} into \cref{eq:ALP-3b,eq:ASP-3b,eq:ATP-3b} and simplify the result with some spinor algebra. 
In $A_{L,\alpha}^{K,us}$, we use 
$\bar u_\nu \slashed{q} P_L v_{\ell_\alpha}  =  - m_{\ell_\alpha}  (\bar u_\nu P_R v_{\ell_\alpha})$, 
and 
 $\bar u_\nu \slashed{P} P_L v_{\ell_\alpha}=  
 2 (\bar u_\nu  \slashed{p}_K P_L   v_{\ell_\alpha}) +   m_{\ell_\alpha} (\bar u_\nu P_R v_{\ell_\alpha})$. 
Similarly, the tensor matrix element $A_{L,\alpha}^{K,us}$ can be rewritten using 
 $p_K^\mu p_\pi^\nu   (\bar u_\nu \sigma_{\mu \nu} P_R  v_{\ell_\alpha}) 
 =  i  (p_K p_\ell - p_K p_\nu)  (\bar u_\nu P_R v_{\ell_\alpha}) 
 + i m_{\ell_\alpha} (\bar u_\nu  \slashed{p}_K P_L   v_{\ell_\alpha})$. 
This leads to 
\begin{align}
\label{eq:kaonamplitudes}
A_{L,\alpha}^{K,us} = A_{R,\alpha}^{K,us}
        &= -\frac{V_{us}}{v^2} \bigg\{
               2 (\bar u_\nu  \slashed{p}_K P_L   v_{\ell_\alpha}) f_+(q^2) \notag\\
        &\hspace{1cm}
             + m_{\ell_\alpha} (\bar u_\nu P_R v_{\ell_\alpha})  \bigg[
                    f_+(q^2) + \big(m_K^2 - m_\pi^2 \big) \frac{f_+(q^2) - f_0(q^2)}{q^2} 
           \bigg] \bigg\} \,,  \\
    A_{S,\alpha}^{K,us}
        &= \frac{V_{us}}{v^2} (\bar u_\nu P_R v_{\ell_\alpha}) 
           \frac{m_K^2 - m_\pi^2}{m_s - m_u} f_0(q^2) \,, \\
    A_{T,\alpha}^{K,us}
        &= \frac{V_{us}}{v^2}
            \Big[ (p_K p_\ell  - p_K p_\nu ) (\bar u_\nu P_R v_{\ell_\alpha})
                + m_{\ell_\alpha} (\bar u_\nu  \slashed{p}_K P_L   v_{\ell_\alpha}) \Big]
            \frac{B_T(q^2)}{m_K} \,.  
\end{align}
Recall that these are amplitudes for $K \to \pi^- \ell_\alpha^+ \nu$. 
For the conjugate process  $\bar K \to \pi^+ \ell_\alpha^- \bar \nu$, proceeding along the same lines we obtain 
\begin{align}
\label{eq:kaonbaramplitudes}
A_{L,\alpha}^{\bar K,us} = A_{R,\alpha}^{\bar K,us}
        &= \frac{V_{us}}{v^2} \bigg\{
               2 (\bar u_{\ell_\alpha}   \slashed{p}_K P_L v_\nu ) f_+(q^2) \notag\\
        &\hspace{1cm}
             - m_{\ell_\alpha} (\bar u_{\ell_\alpha}  P_L v_\nu)  \bigg[
                    f_+(q^2) + \big(m_K^2 - m_\pi^2 \big) \frac{f_+(q^2) - f_0(q^2)}{q^2} 
           \bigg] \bigg\} \,,  \\
A_{S,\alpha}^{\bar K,us}
        &= \frac{V_{us}}{v^2} (\bar u_{\ell_\alpha}  P_L v_\nu ) 
           \frac{m_K^2 - m_\pi^2}{m_s - m_u} f_0(q^2) \,, \\
A_{T,\alpha}^{\bar K,us}
        &=\frac{V_{us}}{v^2}
            \Big[ (p_K p_\nu - p_K p_\ell) (\bar u_{\ell_\alpha}  P_L v_\nu )
        -  m_{\ell_\alpha} (\bar u_{\ell_\alpha} \slashed{p}_K P_L  v_\nu  ) \Big]
            \frac{B_T(q^2)}{m_K} \,.  
\end{align}
To obtain the  amplitudes  $K^+ \to \pi^0 \ell_\alpha^+ \nu$ ($K^- \to \pi^0 \ell_\alpha^- \bar \nu$) one simply multiplies the ones in \cref{eq:kaonamplitudes} (\cref{eq:kaonbaramplitudes}) by the isospin rotation factor $\eta_{K^\pm} = 1/\sqrt 2$.   
The amplitudes for $K_{S,L} \to \pi^- \ell_\alpha^+ \nu$ can be related to \cref{eq:kaonamplitudes,eq:kaonbaramplitudes}   using  
$|K^0_S \rangle = p |K^0 \rangle  - q |\bar K^0 \rangle$,   
$|K^0_L \rangle = q |K^0 \rangle  + p  |\bar K^0 \rangle$, 
where $|p|^2=|q^2| = 1/2$ up to small CP-violating effects.

We next need to evaluate the spin-summed squared production amplitudes, for which we need in particular the spin sums
\begin{align}
    \begin{split}
    \sum_\text{spins} |\bar u_\nu  \slashed{p}_K P_L  v_{\ell_\alpha}|^2
    &= \sum_\text{spins} |\bar u_{\ell_\alpha}  \slashed{p}_K P_L  v_\nu  |^2 = 
    4 (p_K \cdot p_\ell) (p_K \cdot p_\nu) - 2 m_K^2 (p_\ell \cdot p_\nu) \,,  
\\
    \sum_\text{spins} |\bar u_\nu P_R v_{\ell_\alpha}|^2 
    &=  \sum_\text{spins} |\bar u_{\ell_\alpha}  P_L v_\nu |^2  
    =  2 p_\ell \cdot p_\nu \,,     
\\
    \sum_\text{spins} (\bar u_\nu  \slashed{p}_K P_L  v_{\ell_\alpha})  
    (\bar u_\nu P_R v_{\ell_\alpha})
    &=  - \sum_\text{spins} (\bar u_\nu  \slashed{p}_K P_L  v_{\ell_\alpha})
    (\bar u_{\ell_\alpha}  P_L v_\nu) = 
    - 2 m_{\ell_\alpha} (p_K \cdot p_\nu) \,.
    \end{split}
\end{align}
The spin-summed squared amplitudes are then
\begin{align}
 \label{eq:KAON_sumAXY-semileptonic} 
\sum |A_{L,\alpha}^{K_i,us}|^2
        &= \frac{2 \eta_i^2  V_{us}^2}{v^4} \bigg\{  
               4\big [2 (p_K p_\ell) (p_K p_\nu) -  m_K^2 (p_\ell p_\nu)  \big] f_+(q^2)^2 \notag\\
        &\qquad
             - 4 m_{\ell_\alpha}^2 (p_K p_\nu) f_+(q^2) \bigg[ f_+(q^2) + \big(m_K^2 - m_\pi^2 \big)
                                                      \frac{f_+(q^2) - f_0(q^2)}{q^2} \bigg] \notag\\
        &\qquad
            + m_{\ell_\alpha}^2 (p_\ell p_\nu) \bigg[ f_+(q^2) + \big(m_K^2 - m_\pi^2 \big)
                                               \frac{f_+(q^2) - f_0(q^2)}{q^2} \bigg]^2 
                                               \bigg \} \, 
\nnl 
    \sum \bar{A}_{L,\alpha}^{K_i,us} A_{S,\alpha}^{K_i,us}
        &= m_{\ell_\alpha} \frac{2 \eta_i^2   V_{us}^2}{v^4}  \frac{m_K^2 - m_\pi^2}{m_s - m_u} f_0(q^2)  \bigg\{  
               2 (p_K p_\nu) f_+(q^2) \notag\\
        &\qquad
             - (p_\ell p_\nu) \bigg[ f_+(q^2) + \big(m_K^2 - m_\pi^2 \big)
                                                      \frac{f_+(q^2) - f_0(q^2)}{q^2} \bigg] \bigg\} \,,
\nnl 
    \sum \bar{A}_{L,\alpha}^{K_i,us} A_{T,\alpha}^{K_i,us} &= 
    m_{\ell_\alpha} \frac{2 \eta_i^2   V_{us}^2}{v^4} \frac{B_T(q^2)}{m_K} 
    \bigg\{
        -2  \big[ (p_K \cdot p_\ell) (p_K \cdot p_\nu) + (p_K \cdot p_\nu)^2 
        -  m_K^2 (p_\ell \cdot p_\nu) \big] f_+(q^2)
\notag\\ &\qquad
    + \big [  (p_\ell p_\nu) (p_K p_\nu - p_K p_\ell) +  m_{\ell_\alpha}^2 (p_K p_\nu)\big ] 
    \bigg[ f_+(q^2) + \big(m_K^2 - m_\pi^2 \big) \frac{f_+(q^2) - f_0(q^2)}{q^2} \bigg] \bigg\}  \,,
\nnl 
    \sum | A_{S,\alpha}^{K_i,us}|^2
        &= \frac{2 \eta_i^2  V_{us}^2}{v^4} \, (p_\ell p_\nu) 
           \bigg[ \frac{m_K^2 - m_\pi^2}{m_s - m_u} f_0(q^2) \bigg]^2 \,,
\nnl 
    \sum \bar{A}_{S,\alpha}^{K_i,us} A_{T,\alpha}^{K_i,us} &= 
    \frac{2 \eta_i^2  V_{us}^2}{v^4} 
    \frac{m_K^2 - m_\pi^2}{m_s - m_u} f_0(q^2)\frac{B_T(q^2)}{m_K}
    \bigg\{
           (p_\ell p_\nu) (p_K p_\ell - p_K p_\nu)- m_{\ell_\alpha}^2 (p_K p_\nu)
    \bigg\}  \,,
\nnl 
    \sum |A_{T,\alpha}^{K_i,us}|^2
        &= \frac{2 \eta_i^2  V_{us}^2}{v^4} \frac{B_T(q^2)^2}{m_K^2} 
        \bigg\{
                (p_\ell p_\nu) [p_K p_\nu - p_K p_\ell]^2
             + m_{\ell_\alpha}^2 \big[ 2(p_K \cdot p_\nu)^2 -  m_K^2 (p_\ell \cdot p_\nu) \big]
        \bigg\}   \, , 
\end{align}
where
$\eta_i^2 = 1/2$ for $K^\pm$ decays, 
$\eta_i^2 = |p^2| \approx 1/2$ for $K_S$ decays,
and $\eta_i^2 = |q^2| \approx 1/2$ for $K_L$ decays.

%------------------------------------------------------------------
\subsection{Phase Space}
%------------------------------------------------------------------

In order to calculate the production coefficients defined in \cref{eq:pxy}, one needs to decompose the phase space $d \Pi_P$ of the final particles in the production process  into $d \Pi_{P'} d E_\nu$, where $E_\nu$ is the neutrino energy in the lab frame where the target is at rest. 
In this subsection we discuss this decomposition for the leptonic (2-body) $K^+ \to \ell_\alpha^+ \nu$ and semi-leptonic (3-body) $K \to \pi^- \ell^+ \nu$ decays. 

For the 2-body phase space we start from the expression
\beq 
 d \Pi_P = \frac{1}{8 \pi^2 E_\nu} \delta(m_K^2  - m_{\ell_\alpha}^2 
 -2 p_K \cdot k_\nu ) d^3 \mathbf{k}_\nu
=  \frac{E_\nu}{8 \pi^2}  \delta (m_K^2  - m_{\ell_\alpha}^2  - 2 E_\nu E_K  + 2 E_\nu p_K  \cos\theta  )   dE_\nu \, d\cos \theta \, d\phi \,, 
\eeq
where $E_K$ and $E_\nu$ are the kaon and neutrino energies, and $\theta$ and $\phi$ are the polar and azimuthal angle parametrizing the direction of the neutrino momentum with respect to the kaon momentum.  
This expression is valid in any reference frame.
For example, in the kaon rest frame where $E_K= m_K$, $p_K= 0$, $E_\nu \equiv \hat E_\nu$, 
integrating over $\hat E_\nu$ leads to the familiar expression 
$ d \Pi_P = \frac{\hat E_\nu}{16 \pi^2 m_K} d \Omega$ 
with $\hat E_\nu = (m_K^2  - m_{\ell_\alpha}^2)/2$. 
For the present purpose we need to take a different road
so as to decompose $d \Pi_P = d \Pi_{P'} d E_\nu$. 
In the following we assume all kinematic variables are in the lab frame. 
Using the delta function to integrate over $\cos \theta$ we get 
\beq 
d \Pi_P = \frac{1}{16 \pi^2 p_K } d E_\nu  d \phi  . 
\eeq
For unpolarized decays (thus in all practical situations in neutrino experiments)  the spin-summed amplitudes squared are numbers depending only on particles' masses and independent of kinematics. 
Thus we can integrate over $\phi$ to simplify   
\beq 
\label{eq:PROD_PiP2bodyLAB}
 d \Pi_P = \frac{1}{8 \pi p_K } d E_\nu  . 
\eeq
In other words $ d \Pi_{P'} = 1/(8 \pi p_K)$
The integration limits for the neutrino energies are 
$E_\nu \in [E_\nu^{\rm min},E_\nu^{\rm max}] $ with  
\beq
E_\nu^{\rm min} =  \frac{m_K^2 -  m_{\ell_\alpha}^2}{2 (E_K +  p_K )}  ,
\qquad
E_\nu^{\rm max} =  \frac{m_K^2 -  m_{\ell_\alpha}^2}{2 (E_K -  p_K )}  .
\eeq 

For the 3-body phase space it is convenient to start from the parametrization
\begin{align}
d\Pi_P  = \frac{1}{1024 \pi^5 m_K^2} \, dw^2 dq^2
                                        d\!\cos\hat\theta_\nu \, d\hat\phi_\nu \,
                                        d\hat\phi_\pi \,.
\label{eq:dPi3-dPi2-3}
\end{align}
Here, $w^2 = (p_K - p_\nu)^2$, $q^2 = (p_K - p_\pi)^2$, 
$\hat\theta_\nu$ and $\hat\phi_\nu$ are the polar and azimuthal angles of the neutrino momentum in the kaon rest frame, and $\tilde \phi_\pi$ is the azimuthal angle of the pion momentum in the rest frame of the $\pi$--$\ell$ system.
In these variables, the phase space integration limits are 
$\tilde \phi_\pi \in [0,2\pi]$, 
$\hat\phi_\nu \in [0,2\pi]$, 
$\cos \hat\theta_\nu \in [-1,1]$, 
$w^2  \in [(m_\pi+m_{\ell_\alpha})^2, m_K^2 ]$,   
and 
\begin{align}
\label{eq:KAON_qsqminmax}
    q^2_\text{min} &=
        \frac{m_K^2}{2} \bigg[ \frac{w^2 + m_{\ell_\alpha}^2 - m_\pi^2}{w^2}
                             + \frac{m_{\ell_\alpha}^2 + m_\pi^2 - w^2}{m_K^2} 
                             - \bigg( 1 - \frac{w^2}{m_K^2} \bigg) \beta_q \bigg] \,, \nnl
    q^2_\text{max} &=
        \frac{m_K^2}{2} \bigg[ \frac{w^2 + m_{\ell_\alpha}^2 - m_\pi^2}{w^2}
                             + \frac{m_{\ell_\alpha}^2 + m_\pi^2 - w^2}{m_K^2} 
                             + \bigg( 1 - \frac{w^2}{m_K^2} \bigg) \beta_q \bigg] \,, \nnl 
\beta_q & \equiv
        \sqrt{ \bigg(1 - \frac{m_\pi^2 + m_{\ell_\alpha}^2}{w^2} \bigg)^2
                       -  \frac{4 m_\pi^2 m_{\ell_\alpha}^2}{w^4} } 
\, .  
\end{align}
For unpolarized decays, the squared matrix element  summed over spins depends only on $w^2$ and $q^2$.  
If we were interested in the decay width integrated over the neutrino energy we could integrate \cref{eq:dPi3-dPi2-3} over all the angular variables and recover the standard Dalitz form of the 3-body phase space, $d\Pi_P = dw^2 dq^2 / (128 \pi^3 m_K^2)$.
For the present purpose, however, we need to take a different route so as to factor out $d E_\nu$ from  \cref{eq:dPi3-dPi2-3}.  
Indeed, the neutrino energy in the lab frame is a function of $w^2$ and $\cos\hat\theta_\nu$: 
\begin{align}
  E_\nu = \frac{m_K^2 - w^2}{2 m_K^2} \big( E_K + p_K  \cos\hat\theta_\nu \big) \, ,  
\end{align}
where $E_K$ and $p_K$ are the energy and momentum of the kaon in the lab frame. 
It follows that 
\begin{align}
    dw^2 \, d\!\cos\hat\theta_\nu = \frac{2 m_K^2}{p_K (m_K^2 -w^2)} dw^2 \, dE_\nu \,.
\end{align}
Plugging the above into  \cref{eq:dPi3-dPi2-3} and integrating over the azimuthal angles we obtain 
\begin{align}
d\Pi_P =  \frac{1}{128 \pi^3 p_K (m_K^2 -w^2)} dE_\nu \, dw^2 \, dq^2 \,. 
    \label{eq:dPi3-dPi2-5}
\end{align}
The integration limits for the neutrino energy are 
$E_\nu \in  \frac{m_K^2 - w^2}{2 m_K^2}  \big [ E_K - p_K, E_K + p_K \big ]$. 

Furthermore, it is convenient to trade $w^2$ for $\cos \theta$, 
where $\theta$ is the angle between the neutrino and kaon directions in the lab frame.
The reason is that neutrino detectors rarely cover the entire solid angle, and in the new variables it is easier to impose a cut on the neutrino emission angle. 
The variables $w^2$ and $\cos \theta$ are related by 
\beq 
w^2= m_K^2 + 2 E_\nu (p_K \cos \theta - E_K),  
\eeq 
from which it follows that $d E_\nu \, dw^2 = 2 E_\nu p_K \, d E_\nu \, d \cos \theta$. 
Plugging that into \cref{eq:dPi3-dPi2-5} we get 
\begin{align}
d\Pi_P =  \frac{1}{128 \pi^3 (E_K - p_K \cos \theta) } dE_\nu \, d \cos \theta \, dq^2 \,. 
    \label{eq:dPi3-dPi2-6}
\end{align}
The integration limits for $\theta$ and $E_\nu$ are 
$\cos \theta  \in   [\cos \theta_{\rm min} , 1]$,  
$E_\nu  \in  [ 0, E_\nu^{\rm max} ]$, 
where 
\begin{align}
\cos \theta_{\rm min}  &=  {\rm max} \bigg[-1, \frac{2 E_K E_\nu - m_K^2 + (m_\pi + m_{\ell_\alpha})^2}{2 p_K E_\nu}  \bigg] \, ,
\nnl
E_\nu^{\rm max} &= \frac{m_K^2 - (m_\pi + m_{\ell_\alpha})^2}{2 m_K^2 } \big (E_K + p_K) . 
\end{align}

Finally, in experiments such as FASER$\nu$ the kaon energies in the lab frame are not monochromatic, and observables have to be integrated over $E_K$. 
In this case, it is convenient to swap the integration order, so that the $E_K$ integration is performed for a fixed $E_\nu$.  
Then the kaon energy is in the range 
$E_K \in  [  E_K^{\rm min}(E_\nu),  E_K^{\rm max} ]$, 
where 
\beq 
\label{eq:KAON_ekmin}
E_K^{\rm min}(E_\nu)  = \frac{m_K^2}{m_K^2 - (m_\pi+m_\ell)^2}  E_\nu \, ,
\eeq  
and  $E_K^{\rm max}$ is set by the properties of the kaon beam. 
The neutrino energies can be subsequently integrated over the range 
$E_\nu  \in  [ 0, E_\nu^{\rm max}]$ where 
$E_\nu^{\rm max} = \frac{m_K^2 - (m_\pi+m_{\ell_\alpha})^2}{m_K^2} E_K^\text{max}$.

%=======================================================================
\bibliographystyle{JHEP}
\bibliography{nus-v1}
%=======================================================================

\end{document}